\documentclass[trackchanges, twocolumn]{aastex631}

\usepackage{csquotes}
\usepackage{amsmath}
\usepackage{amssymb}
\usepackage{CJK}
\usepackage{booktabs}
\usepackage{enumitem}

\begin{document}
\begin{CJK*}{UTF8}{gbsn}
\title{Lyman-Break Galaxies in the Mpc-Scale Environments Around Three $z\sim 7.5$ Quasars With JWST Imaging}

\correspondingauthor{Maria Pudoka}
\author[0000-0003-4924-5941]{Maria Pudoka}
\affiliation{Steward Observatory, University of Arizona, 933 North Cherry Avenue, Tucson, AZ 85721-0065, USA}
\email{pudoka@arizona.edu}

\author[0000-0002-7633-431X]{Feige Wang}
\affiliation{Department of Astronomy, University of Michigan, 1085 South University Avenue, Ann Arbor, MI 48109, USA}

\author[0000-0003-3310-0131]{Xiaohui Fan}
\affiliation{Steward Observatory, University of Arizona, 933 North Cherry Avenue, Tucson, AZ 85721-0065, USA}

\author[0000-0001-5287-4242]{Jinyi Yang}
\affiliation{Department of Astronomy, University of Michigan, 1085 South University Avenue, Ann Arbor, MI 48109, USA}

\author[0000-0002-6184-9097]{Jaclyn Champagne}
\affiliation{Steward Observatory, University of Arizona, 933 North Cherry Avenue, Tucson, AZ 85721-0065, USA}

\author[0000-0002-2420-5022]{Zijian Zhang}
\affiliation{Kavli Institute for Astronomy and Astrophysics, Peking University, Beijing 100871, People's Republic of China}

\author[0000-0003-2349-9310]{Sof\'ia Rojas-Ruiz}
\affiliation{Department of Physics and Astronomy, University of California, Los Angeles, 430 Portola Plaza, Los Angeles, CA 90095, USA}

\author[0000-0002-2931-7824]{Eduardo Ba\~nados}
\affiliation{Max Planck Institut f\"ur Astronomie, K\"onigstuhl 17, D-69117 Heidelberg, Germany}

\author[0000-0003-4747-4484]{Silvia Belladitta}
\affiliation{Max Planck Institut f\"ur Astronomie, K\"onigstuhl 17, D-69117 Heidelberg, Germany}
\affiliation{INAF - Osservatorio di Astrofisica e Scienza dello Spazio, via Gobetti 93/3, I-40129, Bologna, Italy}

\author[0000-0001-8582-7012]{Sarah E. I. Bosman}
\affiliation{Institute for Theoretical Physics, Heidelberg University, Philosophenweg 12, D–69120, Heidelberg, Germany}
\affiliation{Max Planck Institut f\"ur Astronomie, K\"onigstuhl 17, D-69117 Heidelberg, Germany}

\author[0000-0003-2895-6218]{Anna-Christina Eilers}
\affiliation{Department of Physics, Massachusetts Institute of Technology, Cambridge, MA 02139, USA}
\affiliation{MIT Kavli Institute for Astrophysics and Space Research, Massachusetts Institute of Technology, Cambridge, MA 02139, USA}
\author[0000-0002-5768-738X]{Xiangyu Jin}
\affiliation{Steward Observatory, University of Arizona, 933 North Cherry Avenue, Tucson, AZ 85721-0065, USA}

\author[0000-0003-1470-5901]{Hyunsung D. Jun}
\affiliation{Department of Physics, Northwestern College, 101 7th St SW, Orange City, IA 51041, USA}
\author[0000-0001-6251-649X]{Mingyu Li}
\affiliation{Department of Astronomy, Tsinghua University, Beijing 100084, Peopleʼs Republic of China}

\author[0000-0003-3762-7344]{Weizhe Liu (刘伟哲)}
\affiliation{Steward Observatory, University of Arizona, 933 North Cherry Avenue, Tucson, AZ 85721-0065, USA}

\author[0000-0002-5941-5214]{Chiara Mazzucchelli}
\affiliation{Instituto de Estudios Astrof\'isicos, Facultad de Ingenier\'ia y Ciencias, Universidad Diego Portales, Avenida Ejercito Libertador 441, Santiago, Chile}

\author[0000-0002-4544-8242]{Jan-Torge Schindler}
\affiliation{Hamburger Sternwarte, Universit\"at Hamburg, Gojenbergsweg 112, 21029 Hamburg, Germany}

\author[0000-0003-0643-7935]{Julien Wolf}
\affiliation{Max Planck Institut f\"ur Astronomie, K\"onigstuhl 17, D-69117 Heidelberg, Germany}

\author[0000-0003-0111-8249]{Yunjing Wu}
\affiliation{Department of Astronomy, Tsinghua University, Beijing 100084, Peopleʼs Republic of China}








 
\begin{abstract}

We study the Mpc-scale environments of the three highest redshift luminous quasars at $z\geq 7.5$ (J031343.84-180636.40, J134208.11+092838.61, and J100758.27+211529.21) to understand their connection to large-scale structure.  Cosmological simulations show that these early supermassive black holes (SMBHs) are expected to form in the most massive dark matter halos.  Therefore, it is expected that they are anchors of galaxy overdensities if luminous matter traces the underlying dark matter structure of the Universe. Using JWST NIRCam (F090W/F115W/F250M/F360M/F430M)
imaging, we observe the large-scale structure out to $\sim13$ comoving Mpc around these quasars.  We select F090W-dropout Lyman Break galaxies (LBGs) and F430M-excess [\ion{O}{3}] emitters in the three fields.  We find 18, 21, and 6 
LBG candidates in the fields of J0313, J1342, and J1007, respectively,  resulting in a wide range of overdensities ($1+\delta \sim 19,\,24,$ and $7$).  The photometric redshifts indicate serendipitous foreground and background overdensities in the J0313 field.  The joint angular autocorrelation of the combined LBG sample shows significant clustering on $<1.8$ comoving Mpc scales, demonstrating that the selected galaxies are likely associated with the large-scale structure surrounding the quasars.  This first systematic study of $z\sim 7.5$ quasars shows a diverse set of quasar environments at the onset of their formation, providing empirical data to help constrain theoretical predictions of early structure formation.
\end{abstract}
\section{Introduction}
\end{CJK*}
\label{sec:intro}
Surveys throughout the last twenty years have discovered a substantial population of luminous quasars with black hole masses $\geq 10^{9}\,M_{\odot}$ at $z \gtrsim6$, well into the epoch of reionization \citep[EoR;][]{mortlock, venemans, wu, jiang, wang19, reed, banados23,fan23}.  Furthermore, quasars with black holes of similar masses have been discovered even earlier in cosmic time at $z\sim7.5$, merely 700 Myr after the Big Bang \citep{banados18, yang20, wang21}.  How these supermassive black holes (SMBHs) formed and subsequently grew so soon after the Big Bang has inspired the exploration of different evolutionary tracks for SMBHs.  Many of these evolutionary theories indicate that these SMBHs should reside in overdense environments \citep{overzier09, inayoshi, volonteri}.  

In the $\Lambda$CDM cosmological model, structure begins to form hierarchically when small matter density fluctuations in the early Universe collapse due to gravitational instabilities, creating dark matter halos \citep[DMHs;][]{dodelson, schneider}.  Cosmological simulations based on this concept can produce massive quasars at $z\sim6-7$ by employing particularly high accretion rates (super-Eddington) or seeding the black hole with massive ($>10^{3-4}\,M_{\odot}$) black hole seeds. In these simulations \citep{springel, overzier09}, the high quasar clustering signals \citep{mow, eft}, and quasar abundance matching \citep{lukic} all indicate that these quasars should form within very massive DMHs ($M_{\rm DM}\sim 10^{12-13}M_{\odot}$).

The black holes then grow through accreting cold, low angular momentum gas \citep{dimatteo12} and by hierarchically merging with other black holes \citep{Haehnelt}.  Thus, quasars are expected to reside in environments that are characterized by an overdensity of galaxies given that they 1) reside in the high-$\sigma$ peaks of the Universe's density field, 2) are expected to be surrounded by large gas reservoirs for accretion, and 3) are likely to be proximate to many other black holes with which they will merge.  These overdensities, in extreme cases, can ultimately settle into galaxy clusters around these quasars by $z\sim0$ \citep{angulo}.  In this case, the observed overdensity at high redshift is called a protocluster.

Protoclusters, the most massive structures in the early Universe, reside at the intersections of cosmic filaments that are rich in both dark matter and baryonic gas \citep{overzier16}. The galaxies within them experience a significant inflow of cold gas and frequent mergers making protoclusters the most active regions of galaxy formation during the EoR.  Understanding their formation and their influence on the galaxies within them is instrumental for understanding the large scale structure formation and evolution.  For instance, comparing observed high-redshift protoclusters to cosmological simulations can help to test various theories for dark matter or cosmological structure formation \citep[see][for review]{overzier16}.  Additionally, understanding how the properties of galaxies depend on their environment at early times can shed light on galaxy formation and evolution \citep{nantais, lb}. 

Previous efforts to observe protoclusters in the environments of high-redshift quasars have been limited by the large spatial extent of protoclusters, the faint nature of the galaxies, and the weak Ly$\alpha$ line in the EoR.  Observations of these environments have produced mixed results \citep[e.g.][]{kim09}.  Some authors report galaxy overdensities in the environments of these quasars \citep{kashikawa, utsumi, balmaverde, kashino, wang23} while others find a number of galaxies consistent with a blank field expectation \citep{willot, banados13, mazzucchelli, champagne}.  Additionally, some even report underdense environments \citep{kim09, simpson, goto} or a disparity between the densities measured for Lyman break galaxies (LBGs) versus Ly$\alpha$ emitters \citep[LAEs;][]{ota}.

Multiple explanations for these inconsistencies have been proposed including: investigating too small fields of view (FoVs), strong quasar feedback, and comparing vastly different galaxy populations using a variety of selection techniques.  Many studies utilize Hubble ACS/WFC3 dropout imaging \citep{kim09, simpson, marshall2020a, champagne2023, rojas} or radio interferometers \citep{decarli, champagne, meyer} to identify galaxies at high redshifts.  These instruments have fairly small FoVs (scales of $\sim1-8$ cMpc) which may miss a large portion of the large scale structure considering overdensities at this redshift can extend out to several tens of cMpc \citep{overzier09, chiang13}.  Additionally, it can be difficult to observe LAEs or LBGs in the close vicinity of luminous quasars due to quasar feedback heating the intergalactic medium (IGM) and hampering galaxy formation up to Mpc scales \citep{kashikawa, banados13, chen, lambert}.  Furthermore, inaccurate redshift measurements of the quasar itself can lead to vastly underestimated LAE densities as a result of the Ly$\alpha$ line falling outside of the narrow-band filter \citep{mazzucchelli} while an LBG search would produce different results due to the larger redshift window of broad filters.

Recent JWST NIRCam observations \citep{kashino, wang23, eilers, meyer24} have shown overdensities in these inner regions of high-redshift quasar environments.  This may indicate that the quasar feedback is not strong enough to totally dissociate nearby galaxies, but rather, that these galaxies are simply faint and difficult to detect.  With one NIRCam pointing, \citet{wang23} discovered a filamentary structure containing 10 [\ion{O}{3}]-emitters around the $z=6.6$ quasar J030516.92-315055.9 corresponding to an overdensity of $\delta = 12.6$ within $\sim 9$ arcmin$^2$.  This overdensity persists in a $\sim35$ arcmin$^2$ pointing out to 12 cMpc in \citet{champagne2025a}.  \citet{eilers} observed 4 quasars at $z>6$ with a mosaic of NIRCam WFSS and found a diverse range of densities in their environments.  In particular, they observed the quasar J010013.02+280225.8 (J0100) with an area of $6.5'\times3.4'$.  They found 24 [\ion{O}{3}] emitting systems at the quasar's redshift revealing an overdensity of $\delta\sim 8.2$.  \citet{pudoka} probed an area of $\sim23'\times25'$ around J0100 for LBGs using the Large Binocular Telescope.  They found that the central overdensity observed by \citet{kashino} extends out to cosmic scales nearly a factor of 25 larger than that found by the previous study.  A range of overdensities ($\delta\sim 7-50$) have been reported using JWST observations around $z\sim6-7$ quasars, showing the depth and sensitivity of JWST's NIRCam can easily make up for its restricted FoV and be used to discover the very first protoclusters.  

In this study, we analyze the environments of the three highest redshift quasars known to date (see \S~\ref{sec:fields}) using JWST NIRCam photometry.
With extremely deep imaging of the region around the quasars in the F090W, F115W, F250M, F360M, and F430M filters, we select for LBGs as F090W-dropouts. Additionally, we select line emitters based on F430M flux excess.  The use of medium-band filters helps to constrain these emitters to a narrow redshift range around the quasar redshift.

The structure of this paper is as follows: In \S~\ref{sec:data}, we describe the observations, data reduction, and the photometric catalog used for the subsequent analysis.  \S~\ref{sec:sel} discusses the selection criteria for the LBG candidates and [\ion{O}{3}]-emitters using photometry. We use {\tt EAZY} and {\tt bagpipes} in \S~\ref{sec:SEDs} to calculate the photometric redshifts of the selected galaxies. \S~\ref{sec:results} describes the calculation of the expected number of LBGs in a blank field and the overdensity measurements for each field.  The angular autocorrelation function is calculated in \S~\ref{sec:ACF}.  We discuss the results in \S~\ref{sec:conc} and finally, we summarize our findings and discuss future work in \S~\ref{sec:sum}.  All magnitudes are reported in the AB system and we adopt a $\Lambda$CDM cosmology  ($H_0 = 70\,\mathrm{km\,s^{-1}\,Mpc^{-1}}$, $\Omega_m = 0.3$, $\Omega_{\Lambda} = 0.7$) where $1'\approx2.5$ comoving Mpc (cMpc) at $z=7.5$.  

\section{Data}
\label{sec:data}
\subsection{Quasar Fields}
\label{sec:fields}
We target the environments of the three highest redshift quasars known to date:  J031343.84-180636.40 (J0313) at $z=7.64$ \citep{wang21}, J134208.11+092838.61 (J1342) at $z=7.54$ \citep{banados18}, and J100758.27+211529.21 (J1007) at $z=7.51$ \citep{yang20}.  They all have similar black hole masses derived from the \ion{Mg}{2} emission line of $M_{{\rm BH, J0313}} = (1.6\pm0.4)\times10^{9}\,M_{\odot}$, $M_{{\rm BH, J1342}} = (8.1\pm 1.8)\times10^{8}\,M_{\odot}$, and $M_{{\rm BH, J1007}}=(1.4\pm0.2)\times10^{9}\,M_{\odot}$, respectively \citep{yang21}.   More recently \citet{liu} derived an H$\beta$ black hole mass of $M_{{\rm BH}}=(0.7-2.5)\times10^{9}\,M_{\odot}$ for J1007.  These massive quasars at less than 700 Myr after the Big Bang are ideal regions to search for galaxy overdensities tracing the underlying dark matter structure.

\subsection{JWST Observations}
\label{sec:obs}
The JWST data of the three quasar fields were taken as part of the Cycle 1 GO program \#1764 (PIs: X. Fan, J. Yang, E. Ba\~nados) between 16 December 2022 and 10 May 2023.  We conducted NIRCam imaging using the short wavelength (SW) F090W/F115W filters and long wavelength (LW) F250M/F360M/F430M filters.
The NIRCam
data set covers the dropout filter and the [OIII]+H$\beta$ emission line complex for galaxies in the quasars' environments at $z\sim7.5$.   

For NIRCam imaging, we use an INTRAMODULE three-point dither plus the standard three-point subpixel dither pattern to improve the sampling and to fill the short wavelength detector gaps.  We use the SHALLOW4 readout mode to avoid saturation of the bright central quasars.  Photometric observations with NIRCam are done using filter pairs in the SW and LW channels.  By repeating observations in the F090W filter, we reach a depth in the dropout filter that assures any nondetection is indicative of a strong Lyman break for a clean dropout selection.  This results in total exposure times of 7537 s for F090W, 3768 s for F250M/F360M, and 2802 s for F115W/F430M (Table~\ref{tab:photometry}). 


The NIRCam
imaging was carried out on the FULL array in order to maximize the sky area coverage to properly study the environments of the quasars. Due to the detector gap of NIRCam and the intramodule SW detector gaps, we offset the observations so the quasars do not overlap with any of these gaps while maintaining a relatively large field around the quasar in all directions.  This results in a $\sim0\farcm8$ radius circle of uninterrupted observations centered on each quasar plus an extension of the fields in the southeastern, northern, and southwestern directions from J0313, J1342, and J1007, respectively due to the dimensions of the NIRCam FoV. After masking bright stars, this results in NIRCam image areas of 11.19 arcmin$^2$, 11.23 arcmin$^2$, and 11.24 arcmin$^2$ for the J0313, J1342, and J1007 fields, respectively.  
For a summary of this information, see Table~\ref{tab:photometry}.

As a part of Cycle 1 GO program \#1764, we obtained MIRI imaging in F560W centered on the quasars.  These observations were designed for quasar spectral energy distribution analysis and host galaxy studies.  Due to the small FoV of MIRI compared to the NIRCam observations (4 times smaller) and the shallower depths (3$\sigma \sim 26.5$), this data is insufficient for looking at companion galaxies.  Indeed, of the 47 galaxy candidates introduced in \S~\ref{sec:sel}, only 12 fall in the footprint of the MIRI observations and only 4 of the 12 have reliable ($\sim 2\sigma$) detections.  We therefore do not include this data in the analysis.

\subsection{Data Reduction}
\label{sec:redux}
\subsubsection{NIRCam Imaging}
The NIRCam images were reduced using version 1.10.2 of the JWST Calibration Pipeline \citep[{\tt CALWEBB};][]{jwstpipeline} 
with calibration reference files (\verb|jwst_1080.pmap|) from version 11.16.21 of the Calibration Reference Data System (CRDS).  Beyond the standard {\tt CALWEBB} reduction steps, the data was put through several steps to improve the background noise and astrometry.  These steps are detailed in \citet{wang23} and \citet{yang23} and briefly described below.

Following \cite{schlawin20}, the background-subtracted {\tt rate} files with all detected sources masked were used to iteratively model the {\it 1/f} noise for each amplifier on a column-by-column and row-by-row basis. The results were used as input files for the Stage 2 pipeline.  After the {\it 1/f} noise subtraction and the Stage 2 pipeline,  detector-level noise and sky background remained.  We thus subtracted a master background from the individual Stage 2 output files constructed by combining all exposures observed at similar times.

The astrometric offsets between individual exposures, detectors, and chips were corrected by aligning the images to Gaia DR3.  As there are a limited number of reference stars in individual images, we first created a preliminary mosaic image by processing the LW images for each quasar field through the Stage 3 pipeline.  We aligned this initial mosaic to Gaia DR3 using {\tt tweakwcs} \citep{tweakwcs}.  Then the individual images were aligned to this aligned mosaic image.  The Stage 3 pipeline was used to create drizzled images in each filter with a fixed pixel scale of 0.031\arcsec\ for the SW images and 0.0315\arcsec\ for the LW images with {\tt pixfrac=0.8}.  

The astrometry of the mosaic images was then finalized by further aligning the drizzled images to Gaia DR3, and a final background for each mosaic was modeled and subtracted using {\tt photutils} \citep{photutils}.
\begin{deluxetable}{ccccc}[t!]
    \centering
    \tablecaption{\textbf{Data and Observational Information} The $3\sigma$ depths for each filter, effective area, and total exposure time of the NIRCam imaging data. \label{tab:photometry}}
    \tablehead{\colhead{Filter} & \colhead{J0313} & \colhead{J1342} & \colhead{J1007}  & \colhead{T$_{\rm exp}$ [s]}}
    \startdata
                F090W                 &        29.02         &        29.01         &        28.88                 &        7537           \\ 
                F115W                 &        28.34         &        28.39         &        28.25                 &        2802           \\ 
                F250M                 &        28.68         &        28.68         &        28.45                 &        3768           \\ 
                F360M                 &        29.12         &        29.16         &        28.88                 &        3768           \\ 
                F430M                 &        28.20         &        28.25         &        28.09                 &        2802           \\ \hline
                NIRCam Area           &        11.19         &        11.23         &        11.24                 &                       \enddata
    \tablecomments{AB magnitudes in circular apertures with $d=0\farcs32$. Area in square arcminutes. }
\end{deluxetable}

\subsection{Photometric Catalog Creation}
\label{sec:photom}
We used {\tt SourceXtractor++} \citep{bertin20} on the NIRCam imaging for source detection and photometric catalog creation.  Before extraction, all images were reprojected onto the F360M world coordinate system (WCS) using {\tt reproject} \citep{reproject} and PSF-matched to the resolution of the F360M image using empirical PSF models.  These models for each image were created by stacking bright point sources and Gaia stars detected using {\tt daophot} in the {\tt photutils} package.  

\begin{figure*}[ht]
    \centering
    \includegraphics[width=\linewidth]{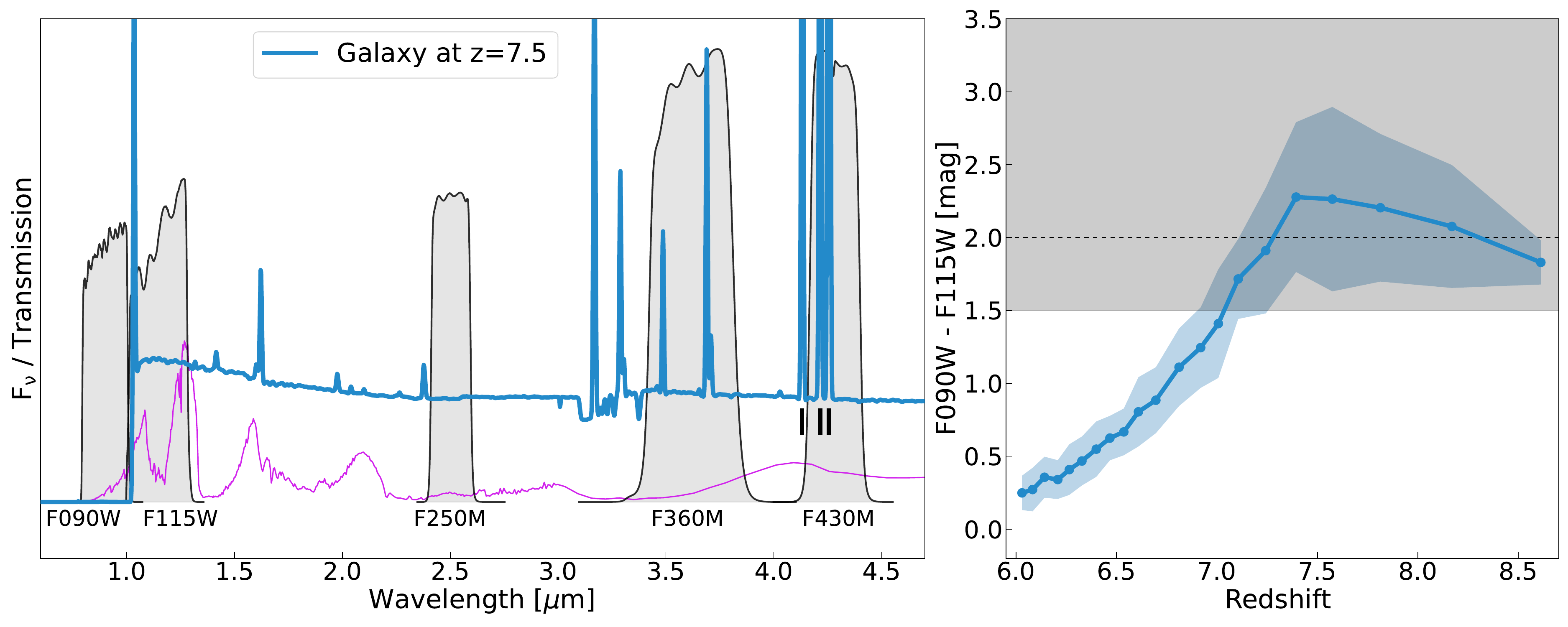}
    \caption{\textit{Left:} Transmission curves of the NIRCam filters used in this analysis are shaded in grey with the  rest-optical spectrum of a galaxy selected from the JAGUAR suite of galaxy templates redshifted to $z=7.5$ is shown in blue. The three black vertical ticks show the H$\beta$ line and [\ion{O}{3}] doublet from left to right.  In magenta is a representative L-type brown dwarf from the Sonora brown dwarf models with $T_{\rm eff} = 1100$ K and (F090W-F115W, F115W-F250M)$=(2.03, -2.41)$.  \textit{Right:} Color vs redshift for the JAGUAR galaxy templates from $z=6$ to $z=9$ calculated using the reported F090W and F115W fluxes in the catalog taking into account the image depths.  Using only galaxy templates that are detected in F115W, the blue shaded region shows the 1$\sigma$ spread in color values in bins containing an equal number of galaxies.  The solid line shows the average value in each bin.  The grey region shows the dropout selection criteria of F090W-F115W$>$1.5 while the black dashed line shows F090W-F115W=2 discussed in the text.}
    \label{fig:dropout}
\end{figure*}

Next, {\tt SourceXtractor++} was used on the PSF-matched images in dual image mode with the F360M filter as the detection image to extract photometry, since F360M is the deepest filter in our data set.  For each filter, the respective {\tt WHT} extension was used as the weight maps. The detection parameters we used that are of note are {\tt DETECT\_THRESH=3.0}, Kron parameters {\tt k, Rmin = 1.2, 1.7}, {\tt DETECT\_MINAREA=10} pixels, and {\tt PHOT\_APERTURES= 12 } pixels.  These specific Kron parameters are chosen to increase the sensitivity to faint and unresolved sources \citep{finkelstein22}.  Due to the smaller aperture sizes, aperture corrections are needed for each source.  To calculate these, we re-ran {\tt SourceXtractor++} with the typical {\tt k, Rmin = 2.5, 3.5} to determine the ratio between the custom and default Kron fluxes in F360M.  These were then applied to every filter as the smaller Kron flux divided by the aperture correction. The photometry in each filter was also corrected for Galactic extinction based on \citet{schlegel98}'s dust reddening map and the dust extinction law derived by \cite{schlafly11}.

We calculated an empirical noise function by placing 1000 empty apertures across each filter's image after masking out detected sources using the segmentation map.  We then measured the 1$\sigma$ spread in flux as a function of aperture size for each of these 1000 apertures ranging from a 1--50 pixel radius. The flux uncertainty for each source was assigned based on the Kron aperture area returned by {\tt SourceXtractor++}.  Additionally, the $3\sigma$ limiting magnitudes values for each field and for each filter is shown in Table~\ref{tab:photometry} for circular apertures with diameter 0.32$''$.

\section{Initial Galaxy Selection}
\label{sec:sel}
\subsection{Lyman Break Galaxies}
Before the end of cosmic reionization, the IGM has a high neutral hydrogen fraction \citep{davies, bosman, spina}.  Due to this abundance of neutral hydrogen, photons with energies higher than the Ly$\alpha$ line will be absorbed causing a sharp decline in the observed flux blueward of the Ly$\alpha$ line at $\lambda_{\rm rest}=1216\,$\AA, referred to as a Gunn-Peterson trough \citep{gunn}.  Taking advantage of this drop in flux which can be seen in Fig.~\ref{fig:dropout} in the left panel, one can efficiently search for star-forming galaxies residing in the EoR using broad-band photometry that covers wide areas without relying on expensive spectroscopy to confirm redshifts. 

At the redshifts of the quasars that we are targeting ($z\approx7.5$), the Lyman break in the rest-frame UV is shifted to the near-infrared (NIR) and falls conveniently between the F090W and F115W filters of JWST/NIRCam.  Therefore, we use the color F090W-F115W as the main property for selecting LBG candidates in our data. We allow galaxies of redshift seven and above to be selected as described below.  We keep the redshift window broad to probe for relatively nearby foreground and background overdensities to be found in the quasars lines of sight.  These types of overdensities have been found in many quasar fields as will be discussed in \S~\ref{sec:results}.

We calculate the typical F090W-F115W colors at $z=6-9$ using v1.2 of the empirically-based mock catalog of star-forming spectra from JAGUAR \citep{jaguar_williams} keeping in mind the possible nondetections based on our image depths. We use all galaxies in the catalog at $6<z<9$ and require them to be detected in the F115W filter. As shown in the right panel of Fig.~\ref{fig:dropout}, the criterion of F090W-F115W $> 2.0$ selects most galaxies at $z\sim7.5$; however, due to nondetections in F090W (which are replaced by the 2$\sigma$ detection limit) many sources will have lower limits in color (i.e. bluer looking colors than they truly are) leading to the range of colors at $z\gtrsim 7.5$ extending below F090W-F115W$=$2. Using this criterion would cause us to only select for the brightest galaxies (in F115W) in our sample.  Therefore, we relax the selection criteria to F090W-F115W $>1.5$ for the initial selection of LBG candidates in order to both select fainter galaxies that may have lower limits in their colors and maintain a large redshift window as mentioned above.

We additionally impose the criterion of having a $3\sigma$ detection in three of the four filters redward of the Lyman break with one of them being the F115W filter.  This assists in removing spurious sources prior to visual inspection while ensuring the galaxy candidate is robust in the F115W filter specifically.  We only allow for the internal {\tt Source-Extractor} flags to be $< 2$ which allows for sources to have a close neighbor but avoids possible extraction errors such as saturation or edge effects.  
\begin{figure}[t!]
    \centering
    \includegraphics[trim={1cm 0.2cm 1cm 1cm}, width=\linewidth, clip]{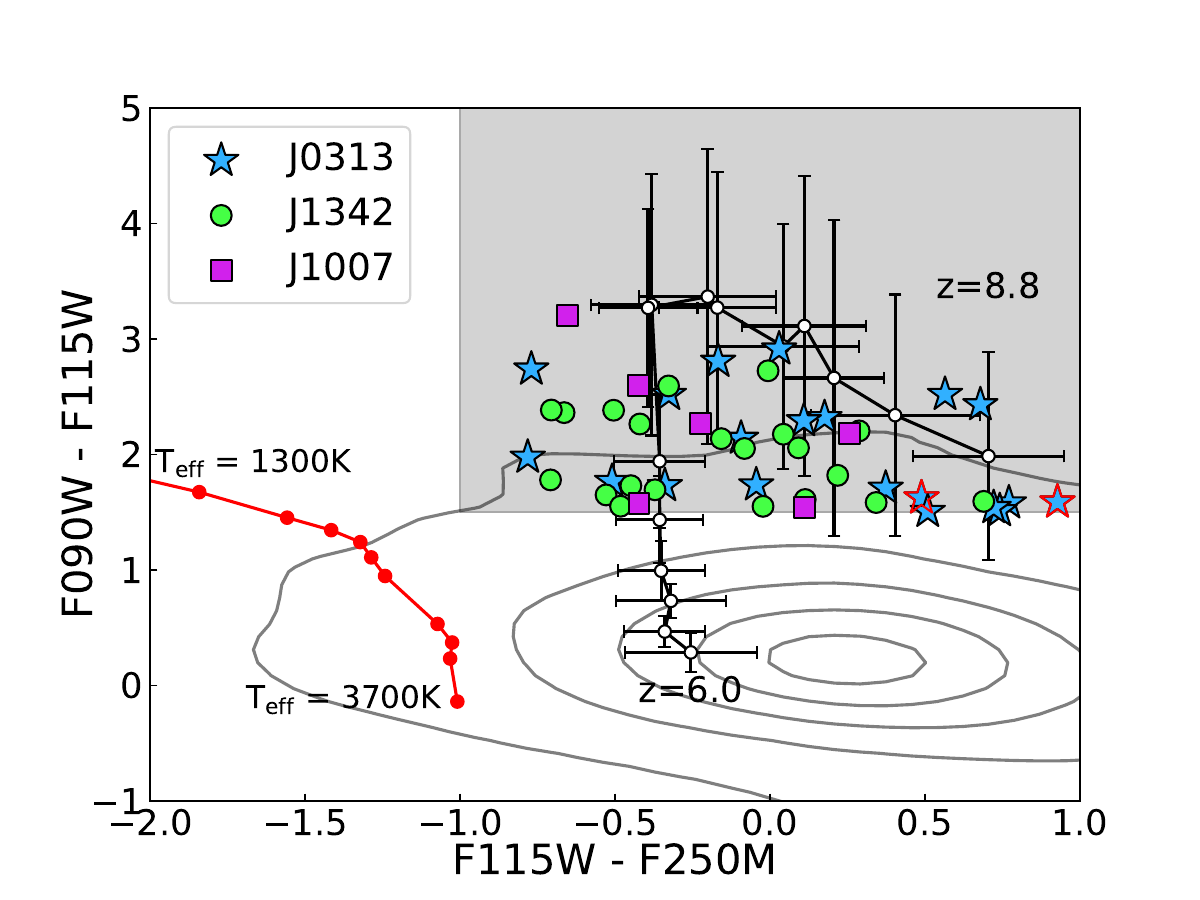}
    \caption{The F090W-F115W versus F115W-F250M color parameter space.  The black points show the track of a typical galaxy from $z=6.0$ to $z=8.8$ calculated from the JAGUAR suite of galaxy templates with errorbars denoting the $1\sigma$ spread in colors for each redshift. The red points show a color track of typical low-mass stars calculated from the Sonora and PHOENIX atmosphere models with $1300 \,{\rm K}\,<T_{\rm eff}<\, 3700$ K. The grey shaded region depicts the color selection criteria. Contours represent the full photometric catalog of the J0313 field (J1342 and J1007 show similar distributions) and colored markers show sources selected by the criteria to be LBG candidates.  The two red outlined stars are sources deemed to be at low redshift from the photometric redshift calculation in \S~\ref{sec:SEDs}.}
    \label{fig:color-color}
\end{figure}

A brown dwarf template with $T_{\rm eff} < 1100$ K from the Sonora brown dwarf atmosphere and evolution models described in \citet{marley} is shown in Fig.~\ref{fig:dropout} in magenta. It can be seen that while the F090W-F115W color will be similar to a high-redshift galaxy, the F115W-F250M color will appear much bluer than a galaxy at high redshift.  Using the models mentioned above and those for K dwarfs from the PHOENIX model atmosphere repository \citep{allard}, we calculate the F090W-F115W and F115W-F250M colors for a suite of models ranging from $T_{\rm eff} = 700$ K to $T_{\rm eff} = 3700$ K with $3.0 \leq \log(g) \leq 5.5$.  This stellar track is shown in Fig.~\ref{fig:color-color} in red. {It is evident that there is a clear deliniation between star and high-redshift galaxy in the parameter space shown in Fig.~\ref{fig:color-color}.  We therefore also include a criterion of $-1<$F115W-F250M$<1$ to avoid the selection of very red low-mass stars.

Additionally, \citet{ryan} estimates that for fields with similar galactic latitudes to these quasar fields, there should be a maximum of 0.17 L to T dwarfs present per square arcminute,  resulting in expecting $\lesssim 5$ in each field. We therefore believe that brown dwarfs are not a significant source of contaminants in our sample. 

After these selection criteria are applied to each quasar field, each candidate goes through visual inspection.  This is a necessary step because spurious sources and image artifacts, such as saturation spikes and noise peaks, are often incorrectly identified as sources by {\tt Source-Extractor} when searching for such faint sources, especially when near the edge of the science image. During this step, a strong visual dropout in the F090W filter along with consistent visual detections in the filters redward of the break are required.  While this is roughly taken care of with the robust selection criteria, there remains some sources in which the photometry is skewed due to bright sources nearby which are thrown out.  We keeping roughly one third of the selected sources in each field.  The majority of removed sources are noise peaks.  The resulting sample of  galaxy candidates will be discussed in \S~\ref{selresults}  and cutouts of each selected source can be seen in the online version of the journal in Figure Set 1 (or in Appendix~\ref{append:cutout} in the arXiv version).  An example of an LBG candidate from the J1342 field is shown in Fig.~\ref{fig:cutout_ex}.

\begin{figure*}
    \centering
    \includegraphics[width=\linewidth]{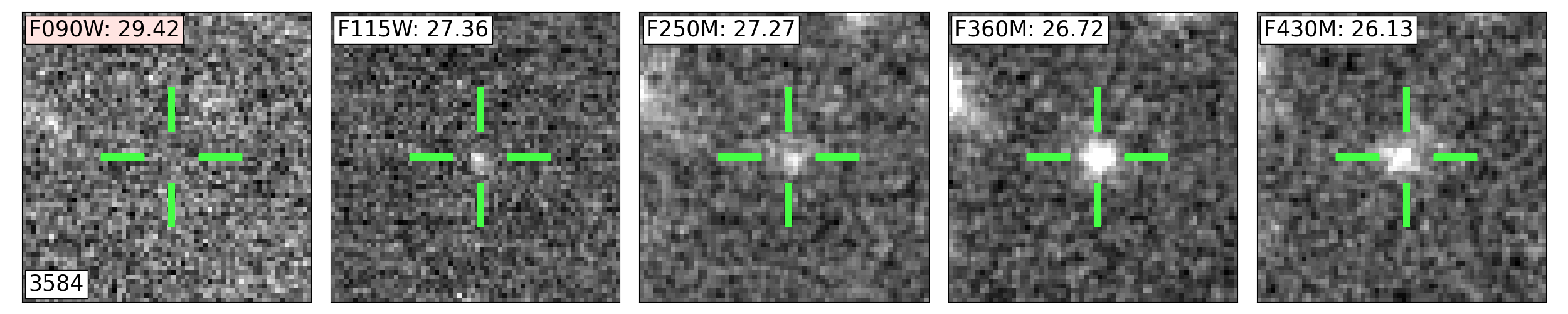}
    \caption{An example of a selected LBG candidate from the J1342 field. Each panel is a $2\times2$ arcminute cutout in the F090W, F115W, F250M, F360M, and F430M filters from left to right with magnitudes shown in the top left of each panel (a red background means it is not detected and replaced by the $2\sigma$ limit. Crosshairs are to guide the eye and each cutout color is normalized separately to account for the large dynamical range of flux in each filter.  The complete figure set (47 images) is available in the online journal (or in Appendix~\ref{append:cutout} in the arXiv version).}
    \label{fig:cutout_ex}
\end{figure*}

\subsection{[\ion{O}{3}]-Emitters}
Typically, [\ion{O}{3}]$\lambda$5007 emission is powered by young stellar populations in a galaxy whose O and B stars ionize regions in the interstellar medium.  These spectral lines have been used to select star-forming galaxies around quasars.  Several recent studies demonstrate that high redshift quasars are embedded in overdensities of these [\ion{O}{3}]-emitters \citep{wang23, eilers, champagne2025a}. Additionally, these lines can arise from the narrow line region of AGN which may make up a fraction of the sources in the fields.  With this in mind, in addition to the Lyman break selection we performed above, we also search for [\ion{O}{3}]-emitters from the sample of LBGs selected above. We note that this is only supplemental and is not used to select against LBG candidates from the previous section.

Similar to the photometric selection of LBGs using a simple broad-band color cut, one can select for [\ion{O}{3}]-emitters with a separate color cut.  As shown in Fig.~\ref{fig:dropout}, the [\ion{O}{3}] doublet falls into the F430M filter at the redshifts of the quasars we are studying.  We can then select for [\ion{O}{3}]-emitters by making a cut on the F250M-F430M color.  The F250M filter should represent the continuum at these redshifts.  We do not choose the F360M filter because it may contain some Balmer and [\ion{O}{3}] lines that could artificially boost the flux in this filter. 

Again, using the JAGUAR galaxy templates we calculate these colors as a function of redshift and show these in the right panel of Fig.~\ref{fig:oiiicolors} for various strengths of the [\ion{O}{3}] equivalent widths (EW).  Strong emitters become very red in F250M-F430M by $z\geq 7.3$ as [\ion{O}{3}]$\lambda$5007 moves into the filter and falls back to bluer colors by $z\geq7.9$ when the [\ion{O}{3}]+H$\beta$ complex moves out of the filter.  It is evident that quiescent galaxies may look red in this color cut due to the Balmer break as shown as the red lines in both panels of Fig.~\ref{fig:oiiicolors}. \citet{Kuruvanthodi} indeed show the Balmer break strengths for a selection of galaxies spanning $z\approx 7-10$ and find that most galaxies at $z\sim7.5$ have flux ratios straddling the Balmer break of $F_{+}/F_{-} \leq 2$.  This corresponds to a magnitude difference of $\Delta_m \approx 0.75$, similar to our mock color calculations.  Therefore, we use this color cut for selecting [\ion{O}{3}]-emitters. 

Based on these colors, we select sources from the LBG candidates with F250M-F430M$>0.75$ and detected at $>3\sigma$ in F430M to ensure strong [\ion{O}{3}] emission.  These criteria allow for the selection of [\ion{O}{3}]-emitters even when the source is not necessarily detected in the F250M filter due to a faint continuum while avoiding contamination from quiescent galaxies.  If the source is not detected in F250M, we set the magnitude to the $2\sigma$ detection limit. Finally, as above, we require the internal {\tt Source-Extractor} flags to be $<2$ for our final selection.

We discuss the selection of [\ion{O}{3}]-emitter candidates next in \S~\ref{selresults}.

\begin{figure*}[ht]
    \centering
    \includegraphics[trim={4cm 0 2cm 2cm}, width=\linewidth, clip]{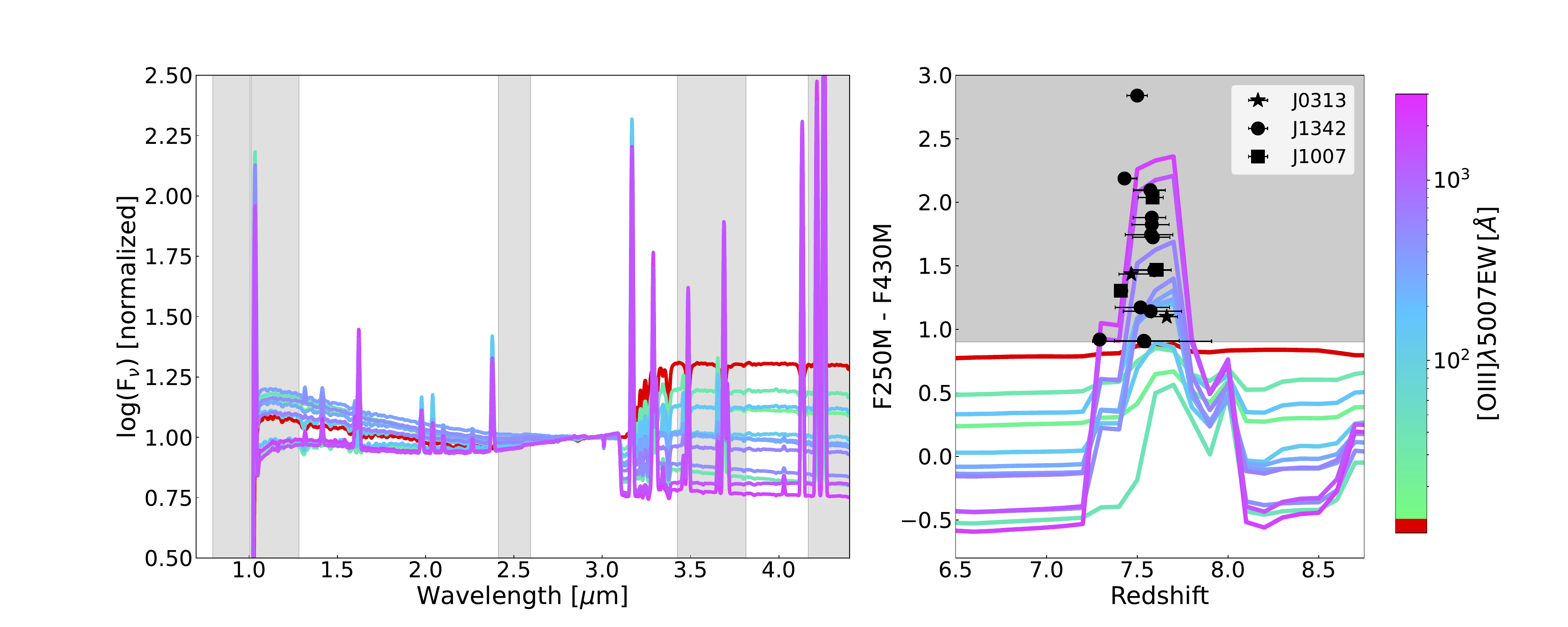}
    \caption{\textit{Left:} A suite of JAGUAR galaxy templates at $z=7.5$ normalized at $\lambda = 3\mu$m with a range of [\ion{O}{3}]$\lambda5007$ equivalent widths from $\log[{\rm EW / \AA}] = 1.3$ (green) to  $\log[{\rm EW / \AA}] = 3.4$ (magenta). The template in red gives an example of a quiescent galaxy with negligible [\ion{O}{3}] emission which can mimic an [\ion{O}{3}]-emitter due to the strong Balmer break. Shaded regions designate the same NIRCam filters as Fig.~\ref{fig:dropout}. \textit{Right:} The F250M-F430M colors of the corresponding galaxy templates as a function of redshift. The grey region denotes the color selection for [\ion{O}{3}] emission in this study.  Note, only strong [\ion{O}{3}]-emitters will be chosen by this selection.  Stars, circles, and squares show the selected [\ion{O}{3}]-emitters in the J0313, J1342, and J1007 fields, respectively with redshift errors shown as the 16th and 84th percentile values from the posterior distributions of the photometric redshift calculation.}
    \label{fig:oiiicolors}
\end{figure*}

\begin{figure*}[ht]
    \centering
    \includegraphics[trim={3.5cm 0.25cm 3cm 1.5cm}, width=0.85\linewidth, clip]{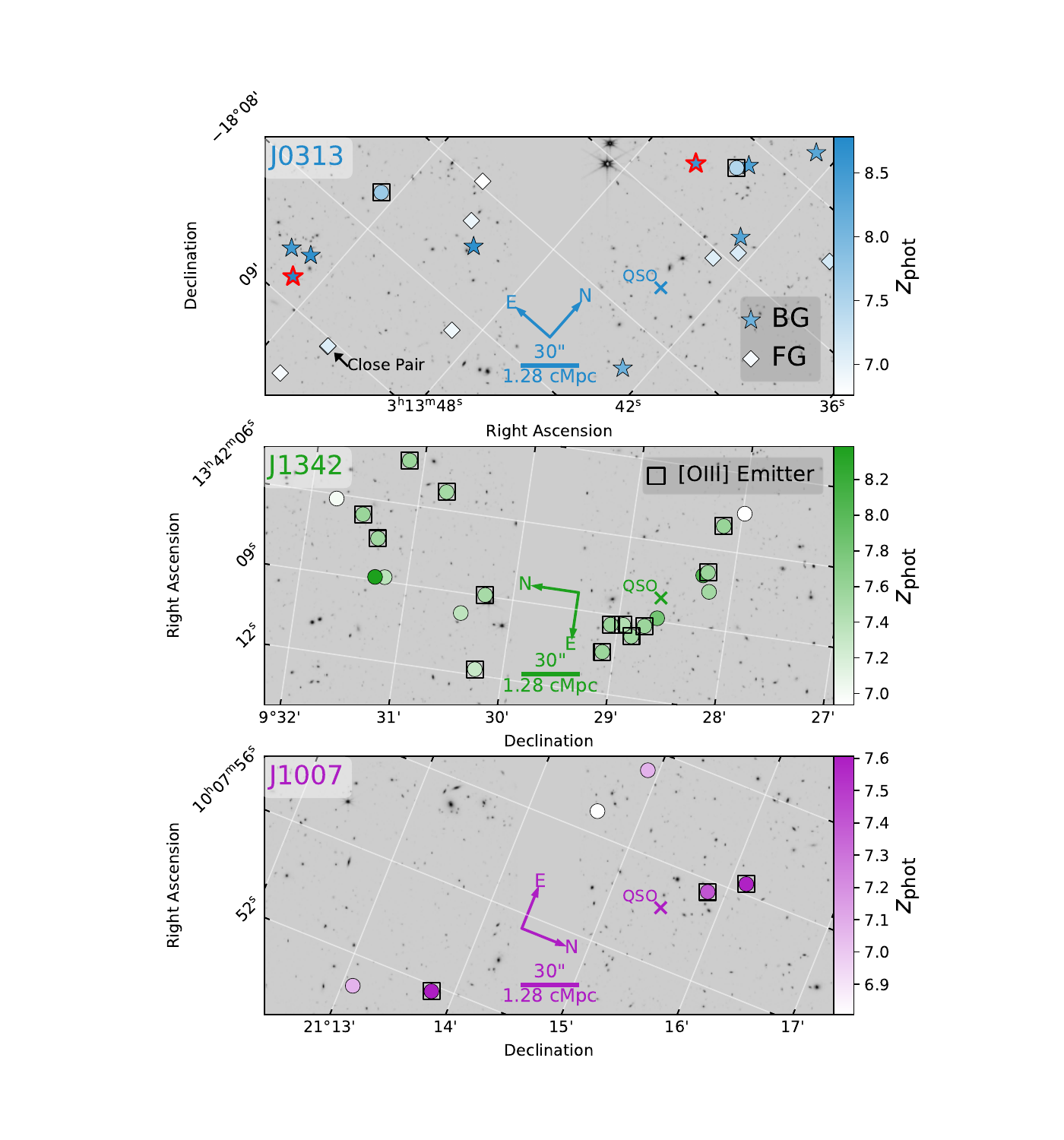}
    \caption{Spatial distribution of candidate LBGs shown as colored markers (according to photometric redshift) superimposed on the F360M image. The quasar location is marked with an x.  Circle markers show candidates at the quasar redshift while, for the J0313 field, stars and diamonds show the BG and FG overdensities, respectively.  Markers surrounded in a box denote [\ion{O}{3}]-emitters while the two red outlined stars show sources removed from the analysis due to photometric redshift solutions discussed in \S~\ref{sec:SEDs}.
    }
    \label{fig:space_dist}
\end{figure*}
\begin{figure*}[ht!]
    \centering
    \includegraphics[trim={3.5cm 0 3.5cm 1cm}, width=\linewidth, clip]{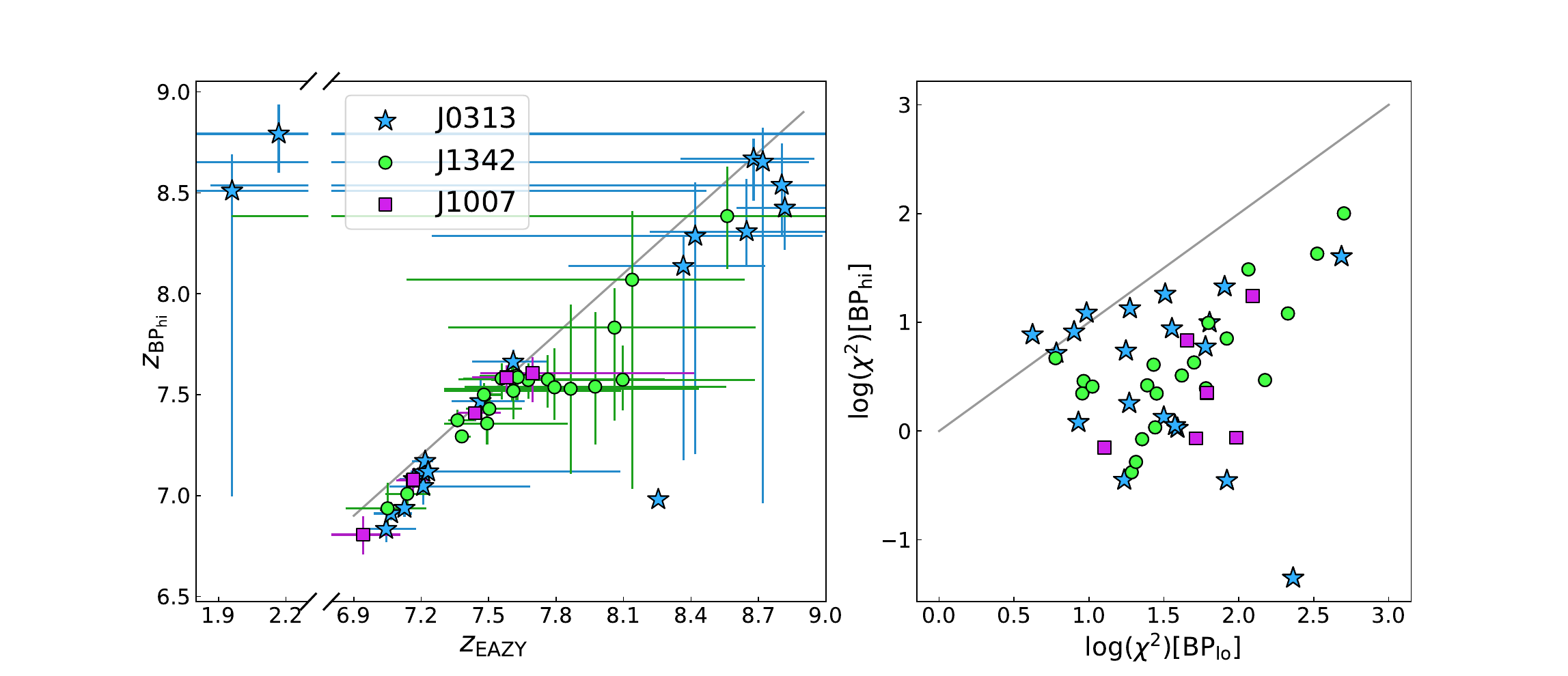}
    \caption{\textit{Left:} Redshift results for the selected LBGs in the three quasar fields with errors reflecting their 16th and 84th percentiles.  This shows the {\tt bagpipes} results constraining $5<z<10$ versus the {\tt EAZY} results.  The grey line denotes a one-to-one relation.  {\tt bagpipes} has a tendency toward lower photometric redshifts compared to {\tt EAZY}, however, most are within the errors. \textit{Right:} The $\chi^2$ values for the {\tt bagpipes} run constraining $5<z<10$ versus the {\tt bagpipes} run constraining $0<z<5$.  The grey line shows the one-to-one relation. Note that the BP$_{\rm hi}$ run has overwhelmingly better fits according to the $\chi^2$ statistic.  Two of the sources with lower BP$_{\rm lo}$ values correspond to the two sources with low-redshift {\tt EAZY} solutions. We remove these sources from the overdensity and clustering analyses below. }
    \label{fig:photoz}
\end{figure*}
\subsection{Galaxy Selection Results}
\label{selresults}
Following the quantitative and visual selections described above, a variety of environments have been revealed around these three highest redshift quasars.  Below are the selection results. The fields with the selected galaxies indicated are shown in Fig.~\ref{fig:space_dist} and the photometric data in the five NIRCam filters is shown in Appendix~\ref{sec:appendprops} in Table~\ref{photo_sel_data}.  

{\bfseries {\scshape J0313:}} Within the J0313 field, we select 20 LBG candidates.  Of these LBG candidates, only two of them are selected as strong [\ion{O}{3}] emitters.  As will be discussed in \S~\ref{sed:results}, this small fraction may be due to these LBGs being at two separate redshifts not associated with the quasar. Additionally, in the same section we do not include sources 4140 and 8403 in our analysis due to low photometric redshift solutions.

{\textbf {\textsc J1342:}} We select 21 LBG candidates within the J1342 field.  16 of these LBGs also show evidence of being [\ion{O}{3}]-emitters, indicating a large fraction of high star-formation rates at the quasar redshift within this field.  

{\textbf {\textsc J1007:}} We select only six LBG candidates within the J1007 field using the same selection criteria as the fields above.  As shown by the calculation of the expected number of LBGs in each field in \S~\ref{sec:results} and Table~\ref{tab:overdensities}, this is not due to the slightly shallower data of the J1007 field.  Of these LBG candidates, three of them are selected as [\ion{O}{3}]-emitters. 

\section{Spectral Energy Distribution Fitting}
\label{sec:SEDs}
In order to further determine if the candidate galaxies are at the quasar redshift, we rely on a robust fitting of the spectral energy distributions (SEDs). While photometric redshifts still have somewhat large systematic uncertainties compared to spectroscopic redshifts, it is evident that this redshift constraint is invaluable compared to simple color selections above.

We use the aperture-corrected Kron photometry described in \S~\ref{sec:photom} to calculate the photometric redshifts of each source using two separate SED fitting programs.  We do this to explore how various fitting techniques and prior assumptions affect the redshift outcomes and galaxy properties and additionally, to make sure we have robust photometric redshifts.  We use the results of the SED fits that use the five NIRCam filters.  

Initially, we use the fast template fitting code {\tt EAZY} to calculate photometric redshifts \citep{brammer}.  This program fits a linear combination of templates with the $\chi^2$ minimization technique.  We use the 16 galaxy templates provided by \citet{hainline} which are optimized to search for high-redshift galaxies and apply the {\tt TEMPLATE\_ERROR.v2.0.zfourge} error template.  For each LBG candidate, we run {\tt EAZY} with a redshift grid between $0<z<12$ with $\Delta z = 0.005$ and no magnitude priors applied.  

We additionally used the Bayesian spectral energy distribution fitting code {\tt bagpipes} \citep{carnall}.  This program relies on the \citet{bandc} stellar population models with a \citet{kroupa} initial mass function and {\tt Cloudy}  nebular emission \citep{ferland} to create complex galaxy models with which to fit the data.  We assume a delayed-$\tau$ model for the star formation history in which the star formation rate goes as $\propto te^{-t/\tau}$.  We allow $\tau$, the timescale of star formation decline, to fluctuate between 1 Myr to 10 Gyr and $t$ to be between 0 and the age of the universe at the given redshift.  We allow the metallicity of the interstellar medium to follow $10^{-6}< Z/Z_{\odot} <10$ and the ionization parameter to be within $-4 < \log U < -2$.  We use a Calzetti dust attenuation curve with $10^{-4} < A_V < 10$ \citep{calzetti}.  All of the above parameters have flat priors. 

We run {\tt bagpipes} twice, the first time (BP$_{\rm lo}$) restricting the redshift to have $0<z<5$ and the second (BP$_{\rm hi}$) to have $5<z<10$.  We do this to compare the $\chi^2 = \sum_i \left(\frac{(F_i-M_i)}{\sigma_i}\right)^2$ values where $F_i(\sigma_i)$ are the observed fluxes(errors) and $M_i$ is the model value for each filter.  This is helpful information when there are two strong peaks in the full redshift probability distribution function (PDF).  Comparing the $\chi^2$ values within each redshift regime helps to determine the best fit for the whole redshift range as it prevents the sampler from falling into a local minimum and not properly sampling the whole parameter space. 

\begin{figure*}[ht!]
    \centering
    \includegraphics[trim={2.5cm 0.5cm 2.5cm 1cm}, width=\linewidth, clip]{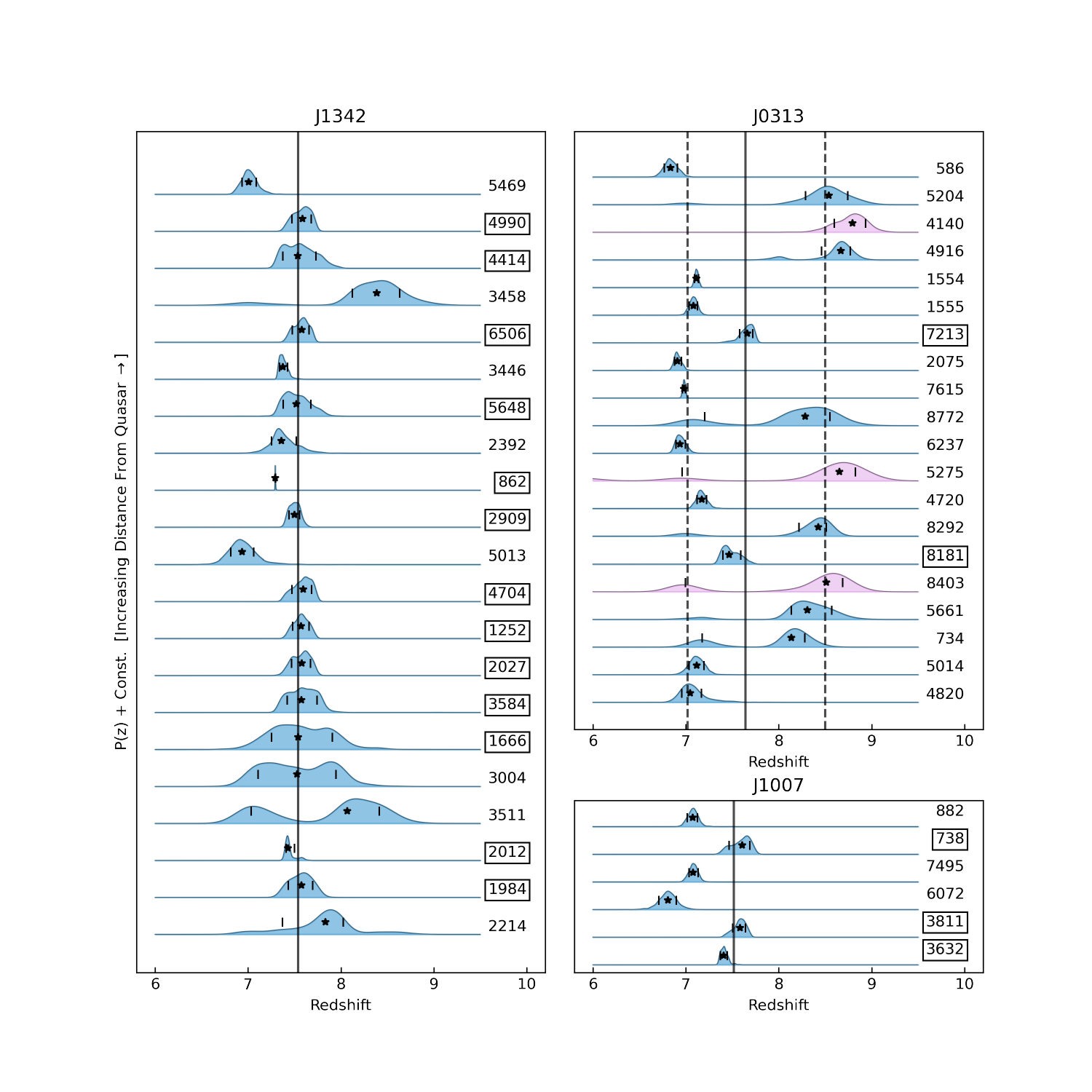}
    \caption{The probability distribution functions of the BP$_{\rm hi}$ photometric redshift calculations normalized to a height of one are shown for each LBG in each field organized by distance to the quasar. Those in magenta are the three sources with $\chi_{\rm lo}^2 < \chi_{\rm hi}^2$. Those on the top/bottom of each respective field are the farthest/closest from/to the quasar. Boxes around the source ID denote candidates selected as [\ion{O}{3}]-emitters.  Stars show the 50th percentile of each PDF and short vertical lines denote the 16th and 84th percentiles.  The solid vertical lines in each panel show the quasar systematic redshift while the dashed lines in the J0313 panel show the average redshift of the FG and BG overdensities.}
    \label{fig:pdfplot}
\end{figure*}

\subsection{Photometric Redshift Results}
\label{sed:results}

The 50th percentile values of the photometric redshift PDFs with the 16th and 84th percentiles as errors for all three runs and their corresponding $\chi^2$ values are reported in Appendix~\ref{sec:appendprops} in Table~\ref{tab:photoz}.  
Sources with [\ion{O}{3}] emission in the F430M filter can constrain the redshift to within $\Delta z\sim0.3$ from the quasar redshift demonstrating the importance of medium-width photometric filters for SED fitting. However, the lack of [\ion{O}{3}] emission should not be interpreted as evidence against the galaxy being at the quasar redshift, as factors such as line strength or obscuration could prevent its detection.

As can be seen in the right panel of Fig.~\ref{fig:photoz}, the $\chi^2$ values for the BP$_{\rm lo}$ run are almost all larger than those of the BP$_{\rm hi}$ run.  There are three sources which have a larger value of $\chi^2$ for the BP$_{\rm hi}$ run (above grey line) and are all in the J0313 field (also see magenta sources in Fig.~\ref{fig:pdfplot}).  Two of these sources (4140 and 8403) correspond to the {\tt EAZY} low-redshift ($z\sim2$) solutions in the left panel of Fig.~\ref{fig:photoz}.  We therefore remove these two sources from the overdensity and clustering analysis below. They are designated in Fig.~\ref{fig:space_dist} by a red outline in the J0313 field and their cutouts can still be found in Figure Set 1 in the online publication.  The remaining source with $\chi^2_{\rm hi}> \chi^2_{\rm lo}$ is 5275.  However, this source has a high-redshift solution from {\tt EAZY} which agrees with the BP$_{\rm hi}$ run and we therefore keep it in the analysis.

The photometric redshift PDFs for each source in each field are shown in the three panels of Fig.~\ref{fig:pdfplot}.  For each source, we plot the PDF with the minimized $\chi^2$ in blue which, as mentioned above, is typically the high-redshift solution.  The three sources in the J0313 field pointed out above show the high redshift solution in magenta. 

It is evident that the three fields show a heterogeneous sample of redshift distributions despite controlling for selection technique and quasar mass/redshift/luminosity. The J1342 and J1007 fields have a redshift distribution centered on the respective quasar redshift within a small range suggesting that these galaxies truly are associated with the large scale structure around the quasar.  However, the J0313 field shows two separate redshift distributions; one in the foreground (FG) with nine galaxies (diamonds in Fig.~\ref{fig:space_dist}), one in the background (BG) of the quasar field with nine galaxies (stars in Fig.~\ref{fig:space_dist}) and only two galaxies at the quasar redshift. 

These FG and BG overdensities are being found more often along the lines of sight to high redshift quasars especially thanks to JWST's ability to get wide field spectroscopic redshifts with NIRCam/WFSS. For example, \citet{wang23}, \citet{yunjingwu}, and \citet{champagne2025a} have discovered and characterized an overdensity around the quasar J0305-3150 with two additional distinct FG overdensities along the line of sight to the quasar. Similarly, \citet{kashino} discovered three overdensities along the line of sight to J0100+2802; one centered on the quasar redshift, one in the FG and one in the BG, while \citet{eilers} reports similar findings for $z\sim 6$ quasars. 

We fit the kernel density estimation of the 50th percentile redshifts for each quasar with a sample of Gaussian distributions in order to characterize each distinct overdensity. The J0313 field can be reproduced with two Gaussian distributions with means $\mu = 7.03, \,8.51$ and standard deviations $\sigma = 0.42, \,0.44$, respectively. The rest of this paper will treat the J0313 field as having two overdensities in the FG and BG while also reporting full-field analysis for completeness. The redshift distributions of the remaining two quasar fields can be reproduced with one Gaussian distribution with $\mu=7.53,\,\sigma=0.19$ and $\mu=7.27,\,\sigma=0.41$ for J1342 and J1007, respectively.  These means are $0.01$ and $0.24$ away from the quasar redshifts, respectively. 
\section{Overdensity Estimates}
\label{sec:results}
\subsection{Completeness Calculations}
\label{compl}

\begin{figure}[ht!]
    \centering
    \includegraphics[trim={0.4cm 0 0.4cm 0cm}, width=\linewidth, clip]{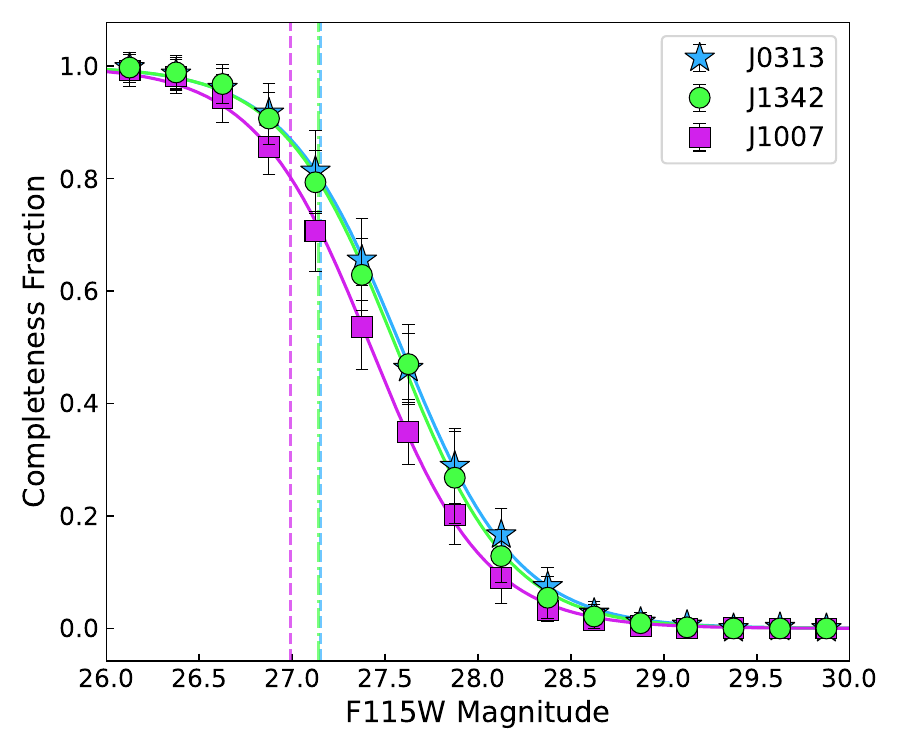}
    \caption{The completeness fraction as a function of apparent magnitude for each field.  Solid lines show the hyperbolic tangent fit to the F115W data while the vertical dashed lines show 80\% completeness at magnitudes 27.15, 27.14, and 26.99 for the J0313, J1342, and J1007 fields, respectively.}
    \label{fig:complete}
\end{figure}

We compute the completeness of each field in the F115W filter using a combination of {\tt Source-Extractor} and {\tt GalSim} \citep{galsim} to simulate and inject galaxies into our science images.  This is needed to compare the galaxy counts to the UV luminosity functions (UVLF) in the literature.  First we run {\tt Source-Extractor} on the F115W science images with the original configuration file in order to produce segmentation maps for each quasar field.  Using both the {\tt WHT} extension and the output segmentation map, we make a boolean mask of the fields.  Then applied to the science image, this creates an object-subtracted science image with only background remaining.  

Using {\tt GalSim}, we simulate galaxies drawing from a uniform distribution of galaxy properties which include effective radius ($R_e \in [1.5, 6]$ arcsec.), S\`ersic index ($n\in[0.5, 4.5]$), ellipticity ($q\in[0.3,1]$), and position angle (PA$\in [0,179]$).  These mock galaxies are then injected into the object-subtracted images avoiding any overlap with the mask created above.  We run {\tt Source-Extractor} on these mock images and crossmatch the results to the input catalog within 0\farcs1. We repeat this procedure 30 times for each observed magnitude bin between 26 and 30 with $\Delta m = 0.25$.  For each run, we produce 50 galaxies within each magnitude bin.  Thus, we inject a total of 1500 mock galaxies for each magnitude bin for a total of 24000 sources in each field.  The resulting completeness curves can be seen in Fig.~\ref{fig:complete} where the errors are a quadrature sum of the Poisson errors and the standard deviation of the 30 trials within each bin.  We find that a hyperbolic tangent function fits the data well for all fields, therefore we run a least-squares fit to the data which can be seen as the solid lines in Fig.~\ref{fig:complete}.  Based on these fits, we find an 80\% completeness rate at magnitudes 27.15, 27.14, and 26.99 for J0313, J1342, and J1007, respectively.  

\subsection{Overdensity Calculation}
We have discovered 20, 24, and 6 LBG candidates in the J0313, J1342, and J1007 fields, respectively. To put these numbers into context, we need an average galaxy density, $\bar{n}$, to which we can compare. Without the information needed to determine a background or average galaxy density from our data, namely spectroscopic redshifts and a large survey volume, we turn to the literature and integrate the galaxy UVLF.  

We make use of a range of UVLFs taking into account that different data sets, data reduction techniques, selection criteria, and cosmic variance may all lead to varying UVLFs.  Employing multiple UVLFs allows for a more robust calculation of the true average expected galaxy number reducing the dependence on a single data set or method.  We do this by fitting a line through the expected galaxy number counts of various UVLFs at $z=7$ and $z=8$ then finding the expected value at the specific redshift of each quasar.  We use the UVLFs from \citet{bouwens21} and \citet{adams24} at $z\sim 8$ and those from 
\citet{bowler15}, 
\citet{atek15}
\citet{bouwens21}, amd 
\citet{harikane24}  at $z\sim 7$.  We convolve these UVLFs with the completeness functions described in \S~\ref{compl}, normalize them to the area of each quasar field, and integrate from $z=7$ to $z=8$.  For each redshift bin, we integrate from the faintest magnitude allowed in F115W given our selection criteria (i.e. $1.5 - 2\sigma[{\mathrm F090W}] = m_{\mathrm F115, \,lim}$).   

In order to get a reliable error on these expected galaxy number counts, we fit each stepwise UVLF reported in the literature separately to a \citet{Schechter} function 
\begin{eqnarray}
    n(M)dM = (0.4\ln10)\Phi^*\left[10^{0.4(M^*-M)}\right]^{\alpha+1} \\\exp\left[-10^{0.4(M^*-M)}\right]dM \nonumber
\end{eqnarray}
with the  Markov chain Monte Carlo (MCMC) method using {\tt emcee} \citep{emcee} with wide uniform priors.  We fit for $\Phi^*, M^*$ and $\alpha$.  From the resulting parameter posteriors, we select 100 random parameter sets and integrate these UVLFs to get a distribution of expected values for each UVLF.  We then use the 16th and 84th percentiles as the errors on the expected value.  Using this process for each of the UVLFs listed above, we fit a line through the calculated expected galaxy count as a function of redshift.  These fits result in a distribution of expected galaxy counts evaluated at the redshift of each quasar.  We use the median as the true expected value and the 16th and 84th percentiles as the errors which are shown in Table~\ref{tab:overdensities}.

\begin{deluxetable}{rcrc}[t!]
    
    \centering
    \tablecaption{\textbf{Overdensity estimate parameters.} The expected number of galaxies in each field, $\bar{n}$, the number of candidates in each field, $n$, and the overdensity estimate.\label{tab:overdensities}}
\tablehead{\colhead{Field} & \colhead{$\bar{n}$\phantom{10}} & \colhead{$n$\phantom{100}} & \colhead{$(1+\delta)$}}
\startdata
J0313\phantom{100}       & $0.95\pm0.04$ \phantom{10}& 18\phantom{100}  & $18.95^{+5.66}_{-4.49}$    \\
J1342\phantom{100}       & $0.87\pm0.04$ \phantom{10}& 21\phantom{100}  & $24.14^{+6.27}_{-5.49}$ \\
J1007\phantom{100}       & $0.90\pm0.03$ \phantom{10}& 6\phantom{100}   & $\phantom{1}6.67_{-2.65}^{+3.98}$  \\ \hline
(FG) J0313\phantom{100}  & $1.29\pm0.05$ \phantom{10}& 9\phantom{100}   & $\phantom{1}6.98_{-2.30}^{+3.19}$  \\
(QSO) J0313\phantom{100} & $0.95\pm0.04$ \phantom{10}& 2\phantom{100}   & $\phantom{1}2.11_{-1.36}^{+2.78}$  \\
(BG) J0313\phantom{100}  & $0.45\pm0.06$ \phantom{10}& 7\phantom{100}   & $15.56^{+8.63}_{-6.10}$ \\
\enddata
\tablecomments{The overdensity for the total J0313 field is only reported for completeness.  The foreground, background, and quasar redshift values should be taken as a more physical calculation of the respective overdensities. (QSO) only takes into account the two candidates near the quasar redshift.}
\end{deluxetable}

To calculate the overdensity, we use the equation 
\begin{equation}
    (1+ \delta) = \frac{n}{\bar{n}}.
    \label{eq:overdense}
\end{equation}  
Where $n$ is the number of galaxies in each field and $\bar{n}$ is the expected number of galaxies calculated with the UVLF.  To calculate the error in overdensity, we propagate $\sigma_{\bar{n}}$ and $\sigma_{n}$ through the equation.  We use the formulae in \citet{Gehrels} to estimate $\sigma_{n}$ due to the small number statistics. This results in overdensties of $1+\delta = 18.95^{+5.66}_{-4.49}, \, 24.14^{+6.27}_{-5.49}$ and $6.67^{+3.98}_{-2.65}$ for J0313, J1342, and J1007, respectively.    We also calculate the overdensities for the separate groups of LBG candidates found in the J0313 field with their photometric redshifts used to calculate the expected galaxy count.  This corresponds to $(1+\delta) = 6.98^{+3.19}_{-2.30},\, 2.11^{+2.78}_{-1.36}$, and $15.56^{+8.63}_{-6.10}$, for the FG, quasar, and BG redshift groups respectively. These values are all reported in Table. \ref{tab:overdensities}.  Values of $1+\delta >1$ indicate densities above the expected field density.  These three quasar fields show a significant diversity in overdensity measurements.

\section{Clustering Properties}
\label{sec:ACF}

\subsection{Angular Autocorrelation Functions}
To quantify the clustering of LBG candidates at $z\sim7.5$, we calculate the angular autocorrelation function (ACF) as the photometric redshifts do not allow us to calculate the full redshift- or real-space ACFs.  This will help validate our selection as we expect galaxies at the same redshift to show hierarchical clustering characterized by a power-law, especially when associated with a possible structural overdensity such as a filament or protocluster.  

The two-point angular ACF calculates the excess likelihood of finding a galaxy at a specific angular distance from another galaxy compared to the expectation from a randomly distributed population \citep{peebles}. If the overdensities in these fields are due to a chance alignment of contaminants, the ACF will not show clustering and the spatial distribution will be consistent with random.  However, if there is truly an overdensity whose galaxies are related spatially, there should be evidence of clustering compared to a random distribution of galaxies.  

We use the \citet{Landy} estimator to calculate the angular ACF which reduces the error to effectively Poisson. This estimator takes the form of 
\begin{equation}
    \omega(\theta) = \frac{\langle DD\rangle - 2 \langle DR \rangle + \langle RR \rangle}{\langle RR \rangle}.
\end{equation}
The values in the above equation are the normalized number of data-data, data-random, and random-random pairs for $\langle DD\rangle$, $\langle DR\rangle$, and $\langle RR\rangle$, respectively.  They take the following form from the raw pair calculations:
\begin{eqnarray}
    \langle DD\rangle =& \frac{2DD}{n_d (n_d -1)}  \nonumber \\
    \langle DR\rangle =& \frac{DR}{n_d n_r} \\
    \langle RR\rangle =&\frac{2 RR}{n_r(n_r-1)} \nonumber
\end{eqnarray}
where $n_d$ and $n_r$ are the total number of selected LBGs and the total number of points in the random catalog, respectively.

It is important to capture the geometry of the field when creating a random distribution catalog as the geometry itself can create a falsely positive clustering signal. We utilize the {\tt WHT} extension as a probability field for each quasar field in order to capture both the depth and geometry, along with the masks used to remove bright/saturated stars. We create $16000$-source random catalogs for each image and transform the xy coordinates to the sky-frame using the WCS. We test the random catalogs by calculating the ACFs of random sources in the photometric catalog and find no clustering signal (open circles in Fig.~\ref{fig:acfplot}), which is to be expected if the random distribution is generated correctly.

Using the data and random catalogs, we use the {\tt corrfunc} \citep{corrfunc} package to count the pairs that go into the ACF calculation.  We set the angular bins logarithmically with widths of $\Delta \log(\theta) \approx 0.3$ out to 4\farcm4 in order to properly probe small scales while not oversampling the larger scales.  There is a close pair in the J0313 field with a separation of 0\farcs4.  To include this, we allow the bins to begin at $\theta \lesssim$ 0\farcs4.

We calculate the clustering of the selected LBG candidates excluding the background overdensity in the J0313 field as these galaxies have $\Delta z\sim1$ from the quasar redshifts while the foreground overdensity only has $\Delta z\sim 0.4$. The resulting ACF is shown in Fig.~\ref{fig:acfplot} with Poisson error bars described in \citet{Landy} assuming an arbitrarily small integral constraint described below.  Extending to the scale of galaxy pairs causes the two bins at $\sim 1$ and $\sim1.5$ arcseconds to not include any galaxy pairs.  This is likely due to a low completeness within the LBG selection considering the trend at $0\farcm05\lesssim\theta\lesssim1\farcm0$ aligns remarkably well with the close pair clustering at $\theta=0\farcs4$. 

\subsection{Power Law Fits}
The typical functional form of the ACF of a clustered population is a power law.  Due to the finite geometry of the field of view, the observed ACF tends toward lower values than the true ACF \citep{groth, roche}.  Thus, the observed ACF can be described in terms of the `true' power law function minus the integral constraint (IC) defined as 
\begin{equation}
    IC = \frac{\sum RR(\theta)\theta^{-\beta}}{\sum RR(\theta)}.
\end{equation}
Therefore, we fit a power law of the form $\omega = A_{\omega}(\theta^{-\beta} - {\rm IC})$ to the ACF. Due to differences in sensitivity between module A and module B of NIRCam, cross-module pairing may introduce power into the ACF at scales $\theta >$2\farcm2.  We therefore exclude bins that include cross-module pairs (the two largest-scale bins).  We also exclude the close pair bin in the fitting procedure.  The fitting range corresponds to the blue solid line in Fig.~\ref{fig:acfplot}.

\begin{figure}[t!]
    \centering
    \includegraphics[trim={0.5cm 0cm 1.6cm 0cm}, width=\linewidth, clip]{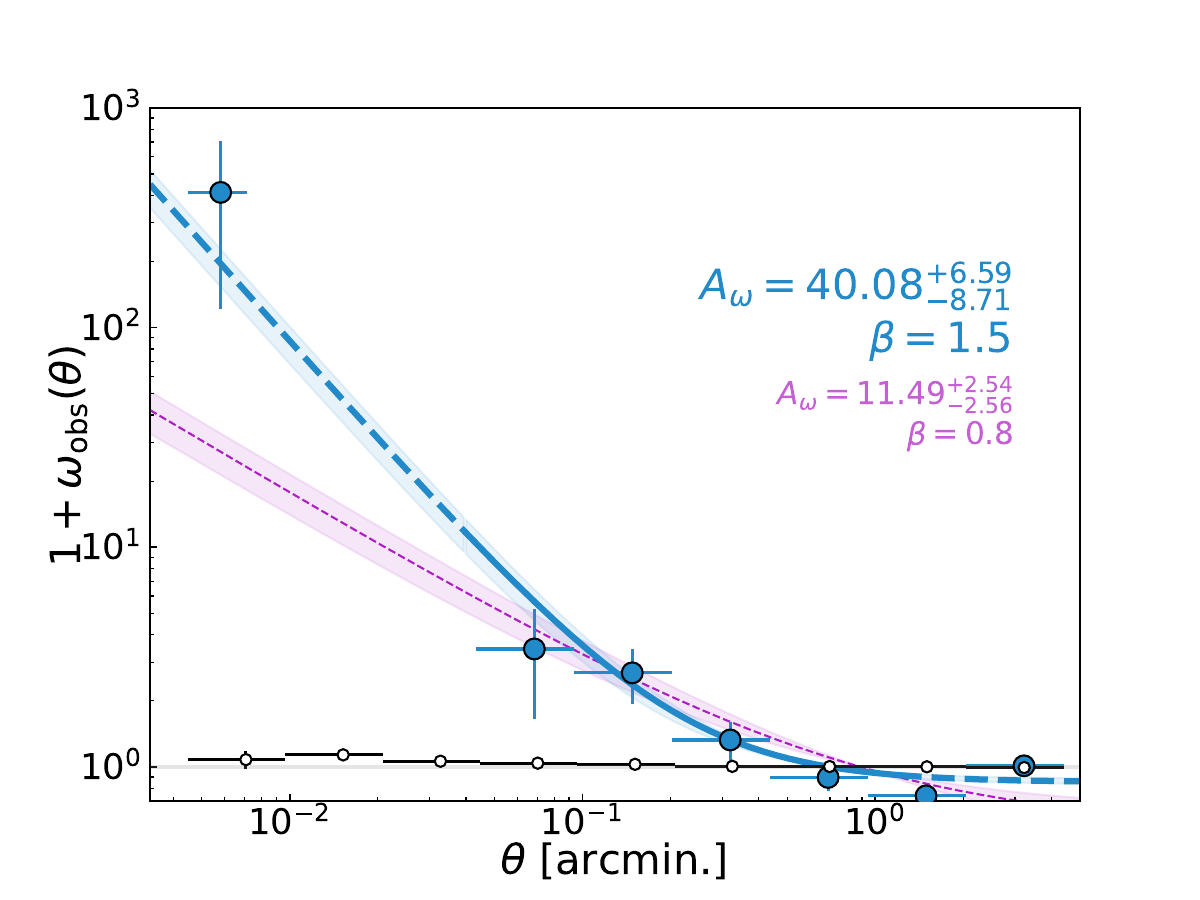}
    \caption{The Landy-Szalay estimated angular ACF for the combined fields shown as blue markers. Errors along the x-axis are the bin sizes while y-axis errors show the Poisson errors described in the text.  Open circles show the ACF for randomly distributed sources within the field consistent with zero.  The solid blue line shows the MCMC fit within the constrained bin range excluding the largest and smallest bins ($\beta=1.5$) while the dashed section shows the fit results extrapolated to the full range of the data.  The shaded regions represents the 16th and 84th percentiles of the posterior distribution of $A_{\omega}$.  In magenta is the same best fit keeping $\beta = 0.8$. }
    \label{fig:acfplot}
\end{figure}

The typical power-law slope for galaxy-galaxy clustering is $\beta=0.8$ \citep{maller,kashino}, though there is some spread in the value of $\beta$ when taken as a parameter to be fit.  For example \citet{garcia} finds a power law slope for the angular ACF of LBGs in quasar fields to be $\beta=1.07\pm 0.49$.  It has also been shown that angular ACFs may be better fit by a double power law function with a steeper slope on small scales \citep{Nikoloudakis}.  To determine which power-law slope is the best fit, we first run an MCMC fitting procedure in which we fit $A_{\omega}$ with a uniform prior and $\beta$ with a Gaussian prior with $\mathcal{N}(\mu=0.8, \sigma=0.7)$.  The resulting posterior distribution of $\beta$ strongly favors a power-law slope of $\beta = 1.5$.  The steeper power-law slope found here is likely due to the field being dominated by the overdense structures within a small FoV.  Without larger scale observations, we keep a single power law fit with a steep slope.

To better constrain the clustering amplitude, we subsequently fit the power law within the range defined above while fixing $\beta=1.5$.  With this fixed slope, $IC = 0.0025$. The resulting clustering amplitude is $A_{\omega} =  40.08^{+6.95}_{-8.71}$ at one arcsecond shown in Fig.~\ref{fig:acfplot}. The dashed line shows the fit extrapolated to the full range of the data demonstrating a remarkable agreement with the smallest bin.  The shaded region in the plot shows the 16th and 84th percentiles in the posterior distribution of $A_{\omega}$.  For consistency with the literature we include the best fit while keeping $\beta=0.8$ in Fig.~\ref{fig:acfplot} (magenta). 

This analysis shows significant clustering on scales up to $25''$ or $\sim 1$ cMpc at $z\sim7.5$.   According to \citet{chiang17}, this clustering is on scales slightly larger than a protocluster core ($\sim 0.2$ cMpc) while being significantly smaller than the average size of a protocluster at $z\sim7$ ($\sim 10$ cMpc). This emphasizes that we are missing the outskirts of the overdensity and that the field is likely dominated by dense large-scale structure rather than an ensemble average of over- and under-dense areas usually probed by very large FoVs.


\section{Discussion}
\label{sec:conc}
These fields host very similar quasars in mass, bolometric luminosity, and redshift.  They also are hosted by galaxies with similar [\ion{C}{2}] gas masses all within $\sim2.4-8.3\times 10^{6}\,M_{\odot}$ \citep{venemans17, wang24}. This would lead one to naively assume that their large-scale environments should be similar as well.  However, despite a homogeneous data set and quasars with comparable nuclear properties, each of these fields displays a significant range of properties both in the fraction of [\ion{O}{3}]-emitters and in the degree of overdensity.  

The quasar J1342 has remarkable evidence of being located within an overdensity of galaxies in addition to it actively undergoing a galaxy merger \citep{banados19}.  This field hosts a highly overdense region ($1+\delta = 24.14^{+6.27}_{-5.49}$).  Additionally, the spatial distribution of the galaxies within this field shows a distinct filamentary structure with many of the galaxies being [\ion{O}{3}]-emitters constraining their redshifts within $\Delta z=0.3$ around the quasar redshift.  If this is truly a filament in the transverse direction of the sky, the quasar resides near the center of the structure supporting the idea that these structures help to funnel gas to the black hole to power the quasar. The large amount of [\ion{O}{3}]-emitters also indicates that this structure may be rich in gas and dust feeding star formation along its path.  

\citet{rojas} searched this quasar field for LBGs using data from HST/ACS and Spitzer finding one LBG at the quasar redshift (C-4636) and two at $z\sim 6.8$ (C-4966 and C-5764).  We recover C-4636 with $z_{\rm phot}=7.57$ (Source ID 1252), but C-5764 is outside of our image footprint and C-4966 is too blue (F090W-F115W $= 0.6$) to be selected as an LBG in our analysis as it is at a lower redshift.  Furthermore, \citet{venemans20} reported a [\ion{C}{2}]-emitter at the quasar redshift as a close companion.  However, it is JWST dark (not detected in NIRCam 
imaging) and it is likely heavily dust obscured.  Therefore, this galaxy has F090W$>30.13$, F115W$>29.48$, F250M$>29.79$, F360M$>30.37$ and F430M$>29.37$.  This demonstrates the importance of using various selection methods to search for diverse galaxy populations as \citet{champagne25b} show distinct galaxy populations may reside in different locations within an overdensity.

The J1007 field shows the least evidence for hosting an overdensity.  Only six galaxies were found in the field resulting in an overdensity of $1+\delta = 6.67^{+3.98}_{-2.65}$.  \citet{schindler} observed this field with the Large Binocular Cameras on the Large Binocular Telecscope (LBT) and JWST/NIRCam photometry with followup NIRSpec/MSA observations and selected one broad-line AGN and 8 surrounding galaxies at $z\sim7.2$.  While the main target of their analysis was the broad-line AGN in the foreground, it is still relevant that some of the lower photometric redshift solutions in the J1007 field in our analysis may be associated with this foreground structure rather than the quasar itself.  Interestingly, we do indeed find two galaxy candidates at $z\sim7.1$ which could be a part of this foreground structure.  If this is the case, the quasar overdensity is even less and closer to the average field density in galaxies.  We note that the broad-line AGN from \citet{schindler} is outside of the footprint of our data.

While the J1342 and J1007 fields show fairly straightforward examples of structure formation and a possible lack thereof, respectively, the J0313 field has ambiguous results regarding the quasar itself hosting an overdensity. Looking at the selected galaxies at face value, it seems that the field hosts an overdensity with $1+\delta = 18.95^{+5.66}_{-4.49}$.  However, a closer analysis of the photometric redshifts indicates that at the quasar redshift, the density is closer to being consistent with the field at $1+\delta = 2.11^{+2.78}_{-1.36}$, even less overdense than J1007. Furthermore, there arise two serendipitous overdensities in the FG and BG. The FG overdensity contains a close galaxy pair with a separation of 0.41$''$ at photometric redshift $z\approx7.1$.  It is not uncommon to find these serendipitous overdensities along the line of sight to high redshift quasars \citep[i.e.][]{wang23, kashino, champagne2025a}; however J0313 is different in that it lacks an overdensity around the quasar itself while those in the literature report line-of-sight overdensities along with a quasar overdensity.

These findings are similar to the results in \citet{eilers} who searched for [\ion{O}{3}]-emitters around four $z\sim6.25$ quasars with JWST/WFSS.  They find a range of overdensities in their analysis from roughly zero to fifty at the scales probed in our analysis (see their Fig. 2).  They find significant overdensities in three of the four quasars observed. Interestingly, only 2 [\ion{O}{3}]-emitters are found at the redshift of J1030$+$0542 while many are found in overdensities along the line of sight in the BG/FG which is similar to the conditions found along the line of sight of J0313.  

At much larger scales ($\sim 60$ cMpc), \citet{morselli} searched for LBGs around $z\sim6$ using the LBT.  They find a range of overdensities from 0.6 to 2.0.  \citet{chiang13} show that the density profiles of high-redshift overdensities decreases significantly with increasing radius.  Following their relation, these overdensities correspond to similar values to our findings at smaller scales.  

These various quasar fields and selection techniques provide evidence for large cosmic variance in which quasars do not necessarily reside in the most overdense areas of the universe at all times.  This may shed light on the various formation histories of the protoclusters themselves rather than the underlying DMHs of the quasars as non-linear baryonic processes begin to play a critical role in structure formation.  It may also point toward different evolutionary pathways for quasar activity and black hole growth such as merger-driven accretion in overdense regions(J1342) versus cold gas accretion or internal processes in the underdense regions (J1007).

\section{Summary and Conclusion}
\label{sec:sum}
\subsection{A Diverse Range of Quasar Environments}
The goal of this paper is to characterize the Mpc-scale environments of the three highest redshift quasars known to date.  According to cosmological simulations, these environments should show evidence of overdensities traced by LBGs.  

We make use of $\sim11$ arcmin$^2$ ($\sim72$ cMpc$^2$) images obtained with JWST's NIRCam in the F090W, F115W, F250M, F360M, and F430M filters
around the the quasars J031343.84-180636.40, J134208.11+092838.61, and J100758.27+211529.21, all at $z\sim7.5$. The depth and sensitivity of NIRCam allow for the selection of LBGs using a photometric dropout selection of F090W-F115W $>$ 1.5.  Additionally, we construct a subsample of the LBGs which show evidence of high [\ion{O}{3}] emission based on the color excess within the F430M filter.  

We calculate the photometric redshifts of each LBG candidate using the five NIRCam filters with both {\tt EAZY} and {\tt bagpipes}.  
The comparison of the results of these fits shows a robust argument that the LBG candidates are indeed at high redshifts.  We measure the completeness limits for each image using simulated galaxies and use this to calculate the expected number of galaxies per field by integrating the UVLF from the literature at the corresponding redshifts.  

Finally, we calculate the galaxy overdensity of each field and measure the clustering of the galaxies using the Landy-Szalay estimator of the angular two-point autocorrelation function.  We fit the ACFs with a power-law function with a steeper power-law slope than is typical ($\beta = 1.5$) possibly indicating overdensity/protocluster core dominance in the FoV. 

We find a significant overdensity at the quasar redshift around J1342 ($1+\delta = 24.14^{+6.27}_{-5.49}$),  while there is little to no overdensity at the quasar redshifts for the J0313 ($1+\delta = 2.11^{+2.78}_{-1.36}$) or J1007 fields ($1+\delta = 6.67^{+3.98}_{-2.65}$).  However, we discover serendipitous FG ($z_{\rm phot}\approx7.1$) and BG ($z_{\rm phot}\approx 8.5$) overdensities in the J0313 field.  

This study reveals an extraordinary diversity in the environments of quasars in the very early Universe.  Despite similar intrinsic properties, one quasar shows a significant overdensity and filamentary structure indicative of early protocluster formation, while others show much less overdense environments.  This heterogeneity challenges the theory that early supermassive black hole growth must be embedded in the most massive dark matter halos resulting in dense environments. However,  this is only if LBGs trace the underlying dark matter distribution with no misalignment due to baryonic physical processes.  Additionally, this could indicate that the expected overdensities have simply not formed yet, providing some insight into the formation timescale of protoclusters.  These findings highlight a need to explore deeper and wider in the early Universe to constrain further the interplay between quasars, baryonic physics, and the large-scale structure. 

\subsection{Future Work}
It is evident that with only four photometric bands plus one dropout band we can only constrain the redshift of these galaxies with a large uncertainty.  Additionally, it has been shown that LBGs and [\ion{O}{3}]-emitters (i.e. highly star forming galaxies) may not populate the same spatial distribution as each other within overdensities \citep{champagne25b}.   A future JWST program, GO \#5221 \citep{pudokajwst}, will target the field of J0313 with a $2\times2$ mosaic of WFSS.  This will not only confirm the existence of the two possible overdensities and their spectroscopic redshifts, but also work to find [\ion{O}{3}]-emitters and their spatial distribution. Additionally, J1342 will be targeted with NIRSpec WFSS in GO \#5911\citep{simcoe}. With this information, we can better characterize the overdensities at larger scales including their kinematics and clustering while ensuring reliable spectroscopic redshifts.

\begin{acknowledgements}
FW acknowledges support from NSF Grant AST-2513040.
JBC acknowledges funding from the JWST Arizona/Steward Postdoc in Early galaxies and Reionization (JASPER) Scholar contract at the University of Arizona.
CM acknowledges support from Fondecyt Iniciacion grant 11240336 and the ANID BASAL project FB210003.
SEIB is supported by the Deutsche Forschungsgemeinschaft (DFG) under Emmy Noether grant number BO 5771/1-1.

This work is based [in part] on observations made with the NASA/ESA/CSA James Webb Space Telescope. The data were obtained from the Mikulski Archive for Space Telescopes at the Space Telescope Science Institute, which is operated by the Association of Universities for Research in Astronomy, Inc. These observations are associated with program \#1764.  Support for program No. 1764 was provided by NASA through a grant from the Space Telescope Science Institute under NASA contract NAS 5-03127.  

This research made use of Photutils, an Astropy package for
detection and photometry of astronomical sources \citep{photutils}.
    
\textit{Software:} NumPy \citep{harris2020array}, Pandas \citep{McKinney_2010, McKinney_2011}, Matplotlib  \citep{Hunter:2007}, Astropy \citep{2018AJ....156..123A, 2013A&A...558A..33A}, SciPy \citep{Virtanen_2020}, EAZY \citep{brammer}, bagpipes \citep{carnall}, Corrfunc \citep{corrfunc}, emcee \citep{emcee}, GalSim \citep{galsim}.
\end{acknowledgements}
\clearpage

\bibliography{refs}{}
\bibliographystyle{aasjournal}

\appendix
\section{LBG Candidate Properties}
\label{sec:appendprops}
The following two tables report the photometric information (Table~\ref{photo_sel_data}) and the photometric redshift information (Table~\ref{tab:photoz}) of the 50 total LBG candidates. 

\startlongtable
\begin{deluxetable*}{lrrrrrrl}

    \centering
    \tablecaption{\textbf{Lyman Break Galaxy Candidate Photometry:} The source ID and aperture corrected flux measurements for each NIRCam filter in nJy for each candidate selected in each field. The [\ion{O}{3}] column denotes whether it is selected as an [\ion{O}{3}]-emitter.\label{photo_sel_data}}
\tablehead{\colhead{Field} & \colhead{Source ID} & \colhead{F090W} & \colhead{F115W}  & \colhead{F250M} & \colhead{F360M} & \colhead{F430M} & \colhead{[\ion{O}{3}]}}
\startdata
J0313 & 586  &  $7.59\pm1.33$ & $38.64\pm2.20$ & $24.17\pm3.13$ & $51.77\pm3.46$ & $37.79\pm\phantom{1}7.64$ & No \\
      & 734  &  $1.81\pm2.45$ & $51.95\pm4.04$ & $61.09\pm4.93$ & $69.27\pm3.08$ & $55.41\pm\phantom{1}6.94$ & No\\
      & 1554 &  $7.05\pm1.25$ & $103.10\pm2.18$ & $105.93\pm2.25$ & $100.50\pm1.63$ & $34.70\pm\phantom{1}5.17$ & No\\
      & 1555 &  $5.76\pm1.59$ & $62.54\pm2.92$ & $46.31\pm3.27$ & $45.04\pm2.35$ & $36.35\pm\phantom{1}6.91$ & No\\
      & 2075 &  $11.94\pm1.66$& $59.10\pm3.07$ & $56.74\pm4.02$ & $51.83\pm2.20$ & $70.51\pm\phantom{1}6.69$ & No\\
      & 4140 &  $9.35\pm3.77$ & $40.36\pm7.05$ & $94.82\pm5.90$ & $91.24\pm3.70$ & $86.67\pm\phantom{1}9.05$ & No\\
      & 4720 &  $3.62\pm2.03$ & $76.51\pm3.39$ & $37.65\pm3.49$ & $51.12\pm2.65$ & $15.03\pm\phantom{1}7.37$ & No\\
      & 4820 &  $3.27\pm1.74$ & $30.20\pm2.92$ & $22.10\pm2.31$ & $22.76\pm1.41$ & $24.62\pm\phantom{1}5.11$ & No\\
      & 4916 &  $3.91\pm4.46$ & $62.53\pm8.19$ & $105.17\pm7.12$ & $111.19\pm4.13$ & $132.86\pm\phantom{1}9.19$ & No\\
      & 5014 &  $5.72\pm3.56$ & $81.42\pm6.22$ & $69.80\pm4.35$ & $62.54\pm3.16$ & $45.49\pm\phantom{1}7.21$ & No\\
      & 5204 &  $4.51\pm2.69$ & $26.32\pm4.78$ & $53.53\pm4.44$ & $52.67\pm2.69$ & $43.21\pm\phantom{1}7.36$ & No\\
      & 5275 &  $5.22\pm1.88$ & $24.74\pm3.38$ & $48.96\pm3.50$ & $38.04\pm2.16$ & $41.25\pm\phantom{1}5.68$ & No\\
      & 5661 &  $-1.95\pm5.65$& $57.56\pm9.12$ & $107.54\pm7.80$ & $155.36\pm4.78$ & $138.77\pm12.36$ & No\\
      & 6237 &  $5.06\pm1.17$ & $29.59\pm2.04$ & $41.72\pm2.29$ & $44.81\pm1.42$ & $47.24\pm\phantom{1}4.46$ & No\\
      & 7213 &  $2.24\pm2.58$ & $50.66\pm4.27$ & $56.02\pm2.99$ & $53.57\pm2.12$ & $154.11\pm\phantom{1}5.58$ & Yes\\
      & 7615 &  $8.16\pm0.47$ & $58.61\pm0.79$ & $53.79\pm0.81$ & $69.94\pm0.68$ & $-17.16\pm\phantom{1}1.91$ & No\\
      & 8181 &  $3.46\pm2.15$ & $37.89\pm3.64$ & $18.45\pm3.09$ & $28.57\pm1.97$ & $69.12\pm\phantom{1}5.54$ & Yes\\
      & 8292 &  $3.10\pm2.02$ & $25.36\pm3.21$ & $49.28\pm2.55$ & $43.44\pm1.75$ & $19.35\pm\phantom{1}4.78$ & No\\
      & 8403 &  $5.52\pm2.16$ & $27.39\pm3.91$ & $42.96\pm2.55$ & $42.29\pm1.84$ & $45.09\pm\phantom{1}5.07$ & No\\
      & 8772 &  $1.99\pm2.63$ & $24.60\pm4.51$ & $39.32\pm3.93$ & $46.41\pm2.58$ & $45.00\pm\phantom{1}7.54$ & No\\
\midrule
J1342 & 862  & $-3.30\pm1.05$ & $54.58\pm\phantom{1}1.87$ & $29.60\pm1.63$ & $18.04\pm1.19$ & $69.09\pm\phantom{1}3.19$ & Yes \\
\phantom{J0313}      & 1252 & $-4.80\pm3.09$ & $76.20\pm\phantom{1}4.87$ & $75.76\pm5.43$ & $88.74\pm3.45$ & $521.94\pm10.52$ & Yes \\
\phantom{J0313}      & 1666 & $-5.46\pm4.93$ & $41.01\pm\phantom{1}7.05$ & $38.02\pm7.08$ & $69.85\pm5.19$ & $87.53\pm15.05$ & Yes \\
\phantom{J0313}      & 1984 & $3.61\pm3.64$ & $28.32\pm\phantom{1}5.50$ & $17.40\pm6.10$ & $41.12\pm3.52$ & $86.79\pm\phantom{1}9.25$ & Yes \\
\phantom{J0313}      & 2012 & $-0.68\pm1.16$ & $31.92\pm\phantom{1}1.99$ & $16.63\pm2.26$ & $19.58\pm1.72$ & $124.68\pm\phantom{1}4.27$ & Yes \\
\phantom{J0313}      & 2027 & $2.46\pm3.93$ & $45.94\pm\phantom{1}5.85$ & $47.80\pm5.26$ & $60.85\pm3.63$ & $256.73\pm\phantom{1}9.59$ & Yes \\
\phantom{J0313}      & 2214 & $3.33\pm3.35$ & $26.65\pm\phantom{1}4.61$ & $36.54\pm4.04$ & $40.44\pm2.99$ & $55.98\pm\phantom{1}8.03$ & No\\
\phantom{J0313}      & 2392 & $-0.69\pm2.88$ & $55.64\pm\phantom{1}4.62$ & $34.96\pm3.87$ & $39.45\pm3.54$ & $52.17\pm\phantom{1}7.50$ & No\\
\phantom{J0313}      & 2909 & $1.43\pm0.78$ & $25.87\pm\phantom{1}1.42$ & $-2.80\pm1.31$ & $11.43\pm0.99$ & $119.10\pm\phantom{1}3.00$ & Yes \\
\phantom{J0313}      & 3004 & $1.24\pm2.01$ & $25.85\pm\phantom{1}3.22$ & $25.32\pm4.04$ & $30.52\pm2.05$ & $35.98\pm\phantom{1}6.83$ & No\\
\phantom{J0313}      & 3446 & $5.83\pm2.09$ & $55.72\pm\phantom{1}3.53$ & $29.09\pm1.85$ & $23.40\pm1.48$ & $52.45\pm\phantom{1}4.04$ & Yes \\
\phantom{J0313}      & 3458 & $3.85\pm4.12$ & $26.90\pm\phantom{1}6.71$ & $50.77\pm5.43$ & $50.53\pm4.24$ & $26.61\pm12.32$ & No\\
 \phantom{J0313}     & 3511 & $3.35\pm3.85$ & $33.10\pm\phantom{1}5.78$ & $40.48\pm3.54$ & $39.21\pm2.71$ & $38.03\pm\phantom{1}7.41$ & No\\
\phantom{J0313}      & 3584 & $1.51\pm5.33$ & $41.21\pm\phantom{1}8.46$ & $44.85\pm6.04$ & $74.58\pm4.21$ & $128.42\pm10.46$ & Yes \\
\phantom{J0313}      & 4414 & $-8.40\pm5.05$ & $67.53\pm\phantom{1}8.75$ & $49.95\pm6.39$ & $79.28\pm4.34$ & $115.46\pm10.87$ & Yes \\
\phantom{J0313}      & 4704 & $7.33\pm6.40$ & $55.83\pm10.41$ & $72.75\pm7.92$ & $83.84\pm5.14$ & $280.74\pm14.22$ & Yes \\
\phantom{J0313}      & 4990 & $4.06\pm1.61$ & $27.30\pm\phantom{1}2.65$ & $30.31\pm2.35$ & $24.47\pm1.69$ & $148.30\pm\phantom{1}4.89$ & Yes \\
\phantom{J0313}      & 5013 & $5.21\pm2.15$ & $29.50\pm\phantom{1}3.34$ & $20.96\pm2.64$ & $28.54\pm1.89$ & $26.81\pm\phantom{1}6.41$ & No\\
\phantom{J0313}      & 5469 & $5.88\pm2.37$ & $44.35\pm\phantom{1}3.74$ & $38.37\pm3.39$ & $30.98\pm2.49$ & $27.61\pm\phantom{1}7.23$& No\\
\phantom{J0313}      & 5648 & $-0.29\pm2.54$ & $30.48\pm\phantom{1}3.79$ & $20.16\pm3.02$ & $36.49\pm2.75$ & $59.36\pm\phantom{1}7.26$ & Yes \\
\phantom{J0313}      & 6506 & $3.08\pm2.36$ & $49.85\pm\phantom{1}3.89$ & $33.86\pm4.08$ & $44.41\pm3.73$ & $191.24\pm\phantom{1}8.53$ & Yes \\
\midrule
J1007 & 738  & $2.50\pm3.38$ & $54.64\pm5.73$ & $69.16\pm6.28$ & $109.19\pm4.32$ & $267.20\pm10.73$ & Yes \\
\phantom{J0313}      & 882  & $2.93\pm0.92$ & $31.27\pm1.46$ & $21.19\pm1.93$ & $31.62\pm1.41$ & $31.99\pm\phantom{1}3.79$ & No\\
\phantom{J0313}      & 3632 & $-1.64\pm4.34$ & $140.02\pm7.50$ & $76.78\pm4.22$ & $77.42\pm3.12$ & $255.06\pm\phantom{1}8.06$ & Yes \\
\phantom{J0313}      & 3811 & $-2.51\pm4.61$ & $59.21\pm7.53$ & $48.16\pm4.66$ & $72.39\pm3.68$ & $314.74\pm\phantom{1}9.24$ & Yes \\
\phantom{J0313}      & 6072 & $16.81\pm4.09$ & $69.32\pm7.01$ & $76.85\pm6.09$ & $110.12\pm4.11$ & $94.78\pm10.15$ & No\\
\phantom{J0313}      & 7495 & $8.26\pm2.33$ & $90.54\pm3.69$ & $61.31\pm3.95$ & $81.18\pm2.93$ & $71.01\pm\phantom{1}6.91$ & No\\
\bottomrule
\enddata
\tablecomments{These values are not corrected for nondetections and therefore are not representative of the $2\sigma$ nondetection values used in the analysis.}
\vspace{-1cm}
\end{deluxetable*}

\startlongtable
\begin{deluxetable*}{llrr|rr|rr|rr}

\tablecaption{\textbf{Photomeric Redshift Information} The source coordinates (hms, dms) and photometric redshifts for the {\tt EAZY}, BP$_{\rm hi}$ and BP$_{\rm lo}$ runs with their 16th and 84th percentile errors are shown with the corresponding $\chi^2$ values for each run. \label{tab:photoz}}
\tablehead{\colhead{\phantom{1}} & \colhead{\phantom{1}}& \colhead{\phantom{1}}& \colhead{\phantom{1}} & \multicolumn{2}{c}{EAZY} & \multicolumn{2}{c}{$\rm BP_{hi}$} & \multicolumn{2}{c}{$\rm BP_{lo}$}} 
\startdata
Field & ID & R.A. & Decl. & z & $\chi^2$ & z & $\chi^2$ & z & $\chi^2$ \\
\midrule
J0313 & 586  & 03:13:52.846 & -18:09:32.813 & $7.04^{+0.13}_{-0.13}$ & 6.11    & $6.83^{+0.08}_{-0.06}$ & 8.74  & $0.84^{+0.03}_{-0.02}$ & 35.76 \\
      & 734  & 03:13:42.879 & -18:07:24.273 & $8.37^{+0.36}_{-0.51}$ & 0.09    & $8.14^{+0.15}_{-0.96}$ & 1.06  & $1.70^{+0.06}_{-0.10}$ & 38.92 \\
      & 1554 & 03:13:52.138 & -18:09:04.160 & $7.21^{+0.02}_{-0.02}$ & 64.65   & $7.11^{+0.02}_{-0.02}$ & 40.20 & $1.71^{+0.01}_{-0.01}$ & 484.13 \\
      & 1555 & 03:13:52.140 & -18:09:03.748 & $7.17^{+0.06}_{-0.06}$ & 0.00    & $7.08^{+0.04}_{-0.04}$ & 0.35  & $1.70^{+0.03}_{-0.02}$ & 83.24 \\
      & 2075 & 03:13:48.897 & -18:08:11.418 & $7.07^{+0.09}_{-0.08}$ & 9.17    & $6.91^{+0.04}_{-0.03}$ & 18.19 & $1.68^{+0.04}_{-0.03}$ & 32.24 \\
      & 4140 & 03:13:54.970 & -18:08:47.557 & $2.17^{+6.96}_{-1.07}$ & 5.23    & $8.79^{+0.14}_{-0.19}$ & 7.69  & $0.54^{+0.12}_{-0.20}$ & 4.20 \\
      & 4720 & 03:13:39.555 & -18:05:23.005 & $7.22^{+0.04}_{-0.06}$ & 20.56   & $7.17^{+0.05}_{-0.05}$ & 21.25 & $1.49^{+0.01}_{-0.01}$ & 80.34 \\
      & 4820 & 03:13:43.071 & -18:06:04.574 & $7.21^{+0.48}_{-0.15}$ & 0.10    & $7.05^{+0.12}_{-0.09}$ & 1.34  & $1.53^{+0.20}_{-0.05}$ & 31.56 \\
      & 4916 & 03:13:54.995 & -18:08:32.128 & $8.68^{+0.27}_{-0.33}$ & 0.76    & $8.67^{+0.10}_{-0.21}$ & 1.80  & $1.56^{+0.14}_{-0.09}$ & 18.56 \\
      & 5014 & 03:13:42.466 & -18:05:53.169 & $7.23^{+0.85}_{-0.08}$ & 0.12    & $7.12^{+0.08}_{-0.08}$ & 1.13  & $1.70^{+0.05}_{-0.04}$ & 37.48 \\
      & 5204 & 03:13:55.748 & -18:08:36.124 & $8.80^{+0.37}_{-6.94}$ & 2.79    & $8.54^{+0.21}_{-0.25}$ & 5.19  & $0.58^{+0.18}_{-0.38}$ & 6.06 \\
      & 5275 & 03:13:50.427 & -18:07:28.201 & $8.72^{+0.20}_{-8.33}$ & 8.68    & $8.65^{+0.17}_{-1.69}$ & 12.18 & $0.37^{+0.14}_{-0.14}$ & 9.64 \\
      & 5661 & 03:13:42.799 & -18:05:45.767 & $8.65^{+0.43}_{-0.43}$ & 0.12    & $8.31^{+0.26}_{-0.17}$ & 0.35  & $0.88^{+0.32}_{-0.11}$ & 17.19 \\
      & 6237 & 03:13:51.161 & -18:07:18.217 & $7.13^{+0.03}_{-0.05}$ & 1.70    & $6.94^{+0.06}_{-0.04}$ & 5.46  & $1.67^{+0.08}_{-0.11}$ & 17.64 \\
      & 7213 & 03:13:54.548 & -18:07:39.675 & $7.61^{+0.15}_{-0.18}$ & 1.93    & $7.66^{+0.06}_{-0.08}$ & 5.95  & $0.28^{+0.01}_{-0.01}$ & 59.87 \\
      & 7615 & 03:13:51.847 & -18:06:57.480 & $8.25^{+0.01}_{-0.01}$ & 1021.42 & $6.98^{+0.01}_{-0.02}$ & 0.04  & $1.50^{+0.00}_{-0.00}$ & 230.26 \\
      & 8181 & 03:13:44.713 & -18:05:18.047 & $7.46^{+0.20}_{-0.13}$ & 3.80    & $7.47^{+0.12}_{-0.07}$ & 9.93  & $0.28^{+0.01}_{-0.01}$ & 63.77 \\
      & 8292 & 03:13:44.413 & -18:05:12.589 & $8.82^{+0.33}_{-0.21}$ & 15.32   & $8.43^{+0.09}_{-0.21}$ & 13.37 & $0.16^{+0.04}_{-0.02}$ & 18.75 \\
      & 8403 & 03:13:46.030 & -18:05:31.265 & $1.96^{+6.51}_{-0.35}$ & 3.17    & $8.51^{+0.18}_{-1.51}$ & 8.19  & $1.58^{+0.17}_{-0.91}$ & 7.95 \\
      & 8772 & 03:13:42.759 & -18:04:42.294 & $8.42^{+0.57}_{-1.17}$ & 0.57    & $8.29^{+0.27}_{-1.08}$ & 1.20  & $1.49^{+0.31}_{-0.73}$ & 8.50 \\
\midrule
J1342 & 862  & 13:42:11.785 & +09:30:15.408 & $7.38^{+0.04}_{-0.03}$ & 57.64  & $7.29^{+0.00}_{-0.00}$ & 100.30 & $1.33^{+0.00}_{-0.00}$ & 502.44 \\
      & 1252 & 13:42:10.441 & +09:29:06.404 & $7.68^{+0.06}_{-0.08}$ & 4.13   & $7.57^{+0.08}_{-0.09}$ & 2.94   & $0.28^{+0.00}_{-0.00}$ & 149.48 \\
      & 1666 & 13:42:09.701 & +09:28:51.679 & $7.97^{+0.58}_{-0.58}$ & 1.23   & $7.54^{+0.37}_{-0.29}$ & 2.63   & $1.56^{+0.48}_{-0.40}$ & 24.47 \\
      & 1984 & 13:42:09.243 & +09:28:45.193 & $7.76^{+0.52}_{-0.40}$ & 0.58   & $7.58^{+0.12}_{-0.14}$ & 4.07   & $1.30^{+2.04}_{-1.02}$ & 26.99 \\
      & 2012 & 13:42:09.313 & +09:28:57.031 & $7.50^{+0.14}_{-0.10}$ & 1.48   & $7.43^{+0.07}_{-0.02}$ & 12.05  & $0.28^{+0.00}_{-0.00}$ & 212.55 \\
      & 2027 & 13:42:09.375 & +09:29:03.973 & $7.60^{+0.13}_{-0.14}$ & 0.38   & $7.58^{+0.09}_{-0.11}$ & 1.08   & $0.28^{+0.01}_{-0.01}$ & 27.63 \\
      & 2214 & 13:42:08.881 & +09:28:38.804 & $8.06^{+0.63}_{-0.74}$ & 0.98   & $7.83^{+0.19}_{-0.46}$ & 2.88   & $1.50^{+0.30}_{-0.24}$ & 9.21 \\
      & 2392 & 13:42:09.763 & +09:30:27.727 & $7.49^{+0.36}_{-0.19}$ & 0.12   & $7.36^{+0.16}_{-0.10}$ & 2.47   & $1.51^{+0.03}_{-0.03}$ & 60.32 \\
      & 2909 & 13:42:08.954 & +09:30:15.700 & $7.48^{+0.08}_{-0.03}$ & 144.39 & $7.50^{+0.06}_{-0.06}$ & 42.73  & $3.08^{+0.07}_{-0.05}$ & 333.28 \\
      & 3004 & 13:42:07.610 & +09:28:12.307 & $7.86^{+0.57}_{-0.56}$ & 0.04   & $7.53^{+0.42}_{-0.42}$ & 0.84   & $1.54^{+0.26}_{-0.20}$ & 22.62 \\
      & 3446 & 13:42:08.840 & +09:31:12.664 & $7.36^{+0.08}_{-0.04}$ & 12.03  & $7.37^{+0.05}_{-0.04}$ & 30.64  & $0.29^{+1.04}_{-0.01}$ & 115.85 \\
      & 3458 & 13:42:08.884 & +09:31:18.013 & $8.56^{+0.55}_{-6.60}$ & 1.70   & $8.38^{+0.25}_{-0.26}$ & 4.68   & $0.64^{+0.29}_{-0.25}$ & 5.98 \\
      & 3511 & 13:42:07.028 & +09:28:16.859 & $8.14^{+0.50}_{-1.00}$ & 0.55   & $8.07^{+0.34}_{-1.04}$ & 2.22   & $1.44^{+0.33}_{-0.83}$ & 9.03 \\
      & 3584 & 13:42:06.896 & +09:28:14.445 & $8.10^{+0.59}_{-0.59}$ & 0.03   & $7.57^{+0.17}_{-0.15}$ & 0.42   & $3.19^{+0.35}_{-1.91}$ & 19.27 \\
      & 4414 & 13:42:07.430 & +09:31:19.490 & $7.79^{+0.44}_{-0.39}$ & 2.81   & $7.54^{+0.19}_{-0.16}$ & 4.27   & $1.50^{+0.33}_{-0.18}$ & 50.21 \\
      & 4704 & 13:42:05.076 & +09:28:09.481 & $7.62^{+0.12}_{-0.16}$ & 1.47   & $7.59^{+0.09}_{-0.12}$ & 2.57   & $0.28^{+0.01}_{-0.01}$ & 10.54 \\
      & 4990 & 13:42:06.614 & +09:31:29.741 & $7.63^{+0.10}_{-0.14}$ & 12.23  & $7.59^{+0.09}_{-0.11}$ & 9.88   & $0.29^{+0.01}_{-0.00}$ & 62.65 \\
      & 5013 & 13:42:04.497 & +09:27:58.910 & $7.05^{+0.17}_{-0.19}$ & 0.02   & $6.94^{+0.13}_{-0.12}$ & 0.52   & $1.49^{+0.38}_{-0.64}$ & 20.58 \\
      & 5469 & 13:42:06.167 & +09:31:45.561 & $7.14^{+0.10}_{-0.10}$ & 0.73   & $7.01^{+0.08}_{-0.07}$ & 2.21   & $1.64^{+0.08}_{-0.11}$ & 28.26 \\
      & 5648 & 13:42:05.321 & +09:30:45.380 & $7.61^{+0.48}_{-0.31}$ & 0.03   & $7.52^{+0.16}_{-0.14}$ & 3.24   & $1.51^{+0.57}_{-0.06}$ & 41.63 \\
      & 6506 & 13:42:04.347 & +09:31:08.290 & $7.56^{+0.18}_{-0.17}$ & 0.81   & $7.58^{+0.08}_{-0.10}$ & 7.09   & $0.28^{+0.00}_{-0.01}$ & 83.03 \\
\midrule
J1007 & 738  & 10:07:51.749 & -21:13:47.813 & $7.70^{+0.72}_{-0.23}$ & 0.53 & $7.61^{+0.08}_{-0.14}$ & 0.85 & $0.26^{+0.01}_{-0.00}$ & 51.74 \\
      & 882  & 10:07:50.782 & -21:13:05.907 & $7.16^{+0.07}_{-0.07}$ & 0.30 & $7.07^{+0.05}_{-0.06}$ & 2.25 & $1.50^{+0.01}_{-0.01}$ & 61.06 \\
      & 3632 & 10:07:59.546 & -21:15:50.297 & $7.44^{+0.11}_{-0.08}$ & 1.69 & $7.41^{+0.04}_{-0.04}$ & 17.41& $1.23^{+0.00}_{-0.00}$ & 123.84 \\
      & 3811 & 10:08:00.410 & -21:16:08.665 & $7.58^{+0.15}_{-0.15}$ & 0.30 & $7.59^{+0.06}_{-0.08}$ & 6.84 & $0.28^{+0.01}_{-0.01}$ & 45.26 \\
      & 6072 & 10:08:00.897 & -21:14:36.379 & $6.94^{+0.17}_{-0.18}$ & 0.05 & $6.81^{+0.09}_{-0.10}$ & 0.71 & $0.84^{+0.45}_{-0.05}$ & 12.65 \\
      & 7495 & 10:08:03.164 & -21:14:53.978 & $7.16^{+0.06}_{-0.07}$ & 0.59 & $7.08^{+0.06}_{-0.05}$ & 0.87 & $1.50^{+0.01}_{-0.01}$ & 96.06 \\
\bottomrule
\enddata
\tablecomments{Errors shown as $0.00$ are due to very narrow PDFs for which the percentile differenes are $<0.01$.}
\end{deluxetable*}

\section{LBG Cutouts}
\label{append:cutout}
The $2\times2$ square arcminute cutouts for all 47 LBG candidates are shown below. 

\begin{figure}
\centering
\includegraphics[width=\linewidth]{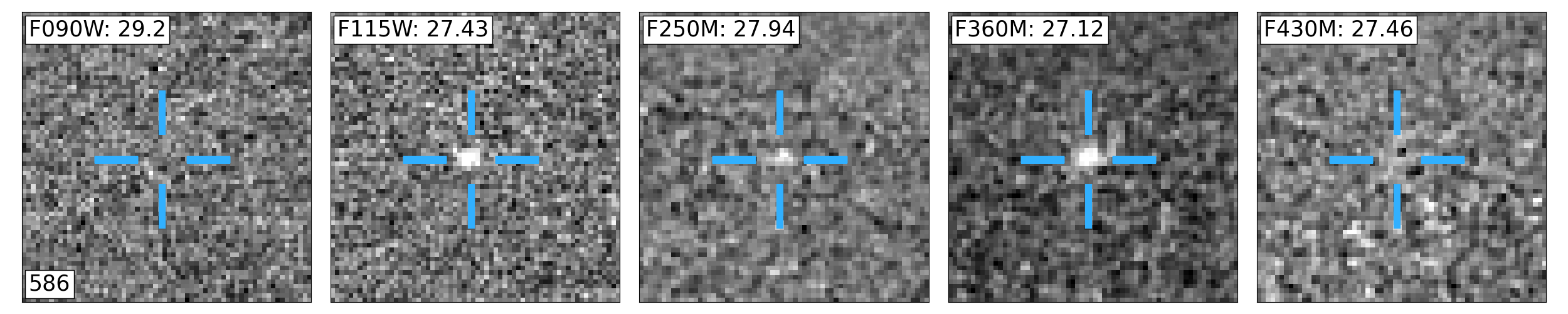}
\includegraphics[width=\linewidth]{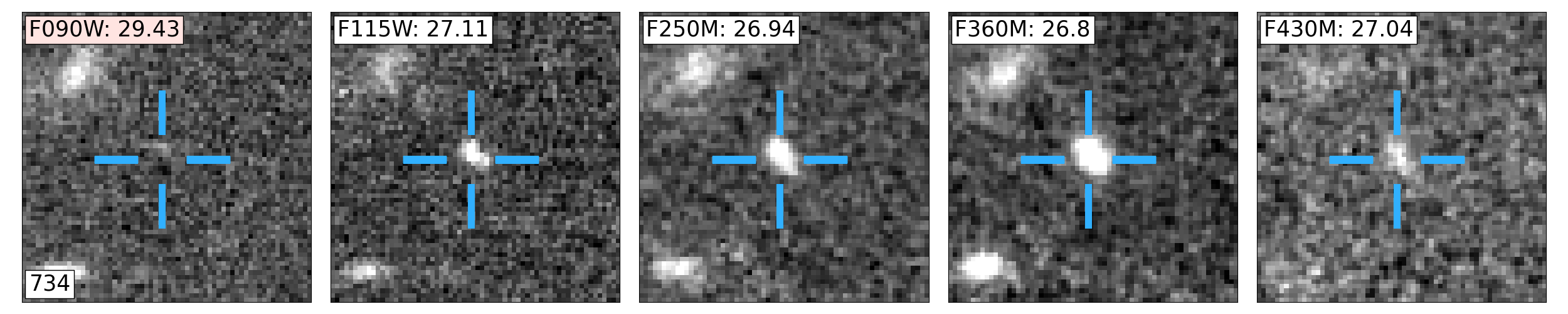}
\includegraphics[width=\linewidth]{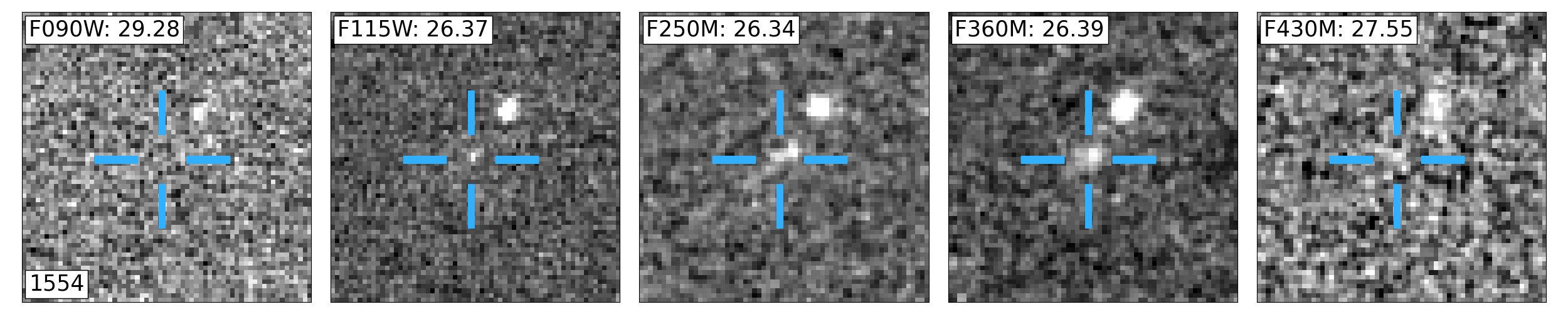}
\includegraphics[width=\linewidth]{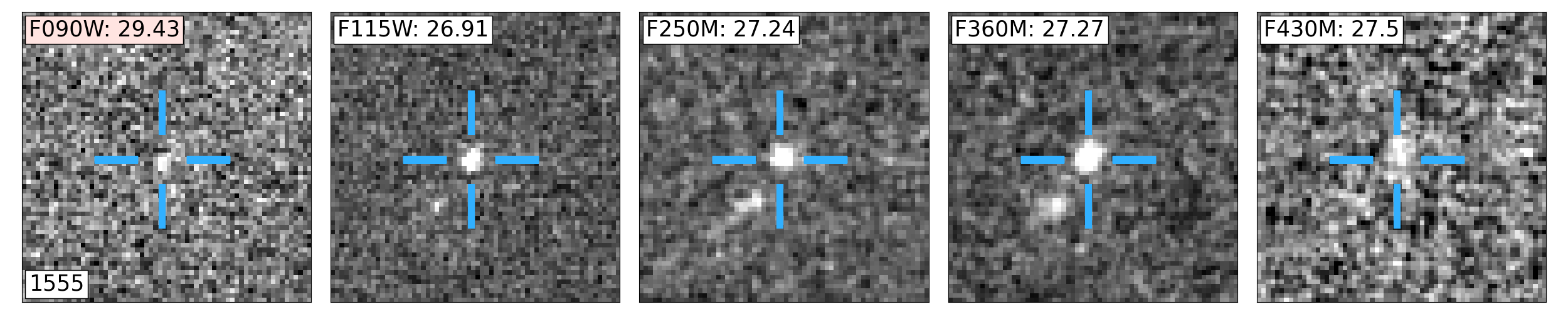}
\includegraphics[width=\linewidth]{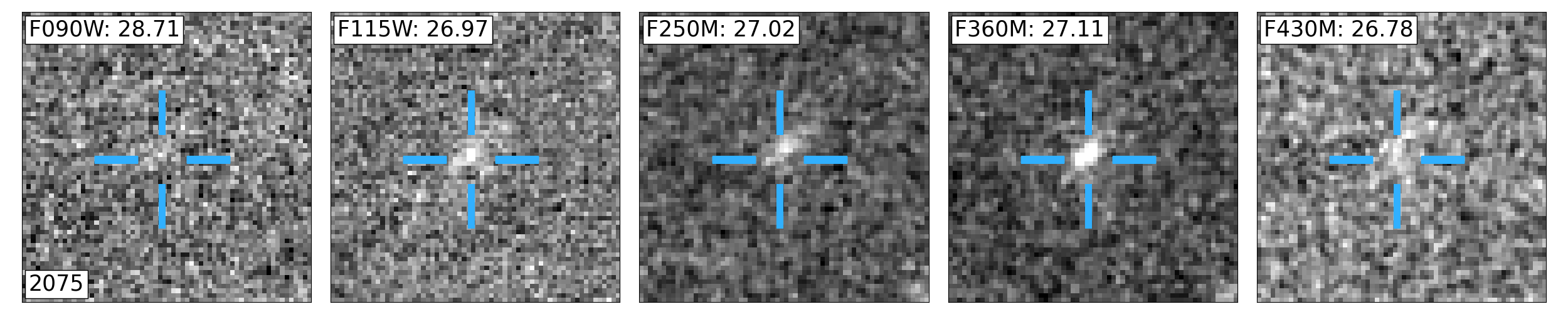}
\caption{Cutouts of each selected LBG candidate. Each panel is a $2\times2$ arcminute cutout in the F090W, F115W, F250M, F360M, and F430M filters from left to right with magnitudes shown in the top left of each panel (a red background means it is not detected and replaced by the $2\sigma$ limit). Crosshairs are to guide the eye and each cutout color is normalized separately to account for the large dynamical range of flux in each filter.}
\label{fig:cut1}
\end{figure}

\begin{figure}
\centering
\includegraphics[width=\linewidth]{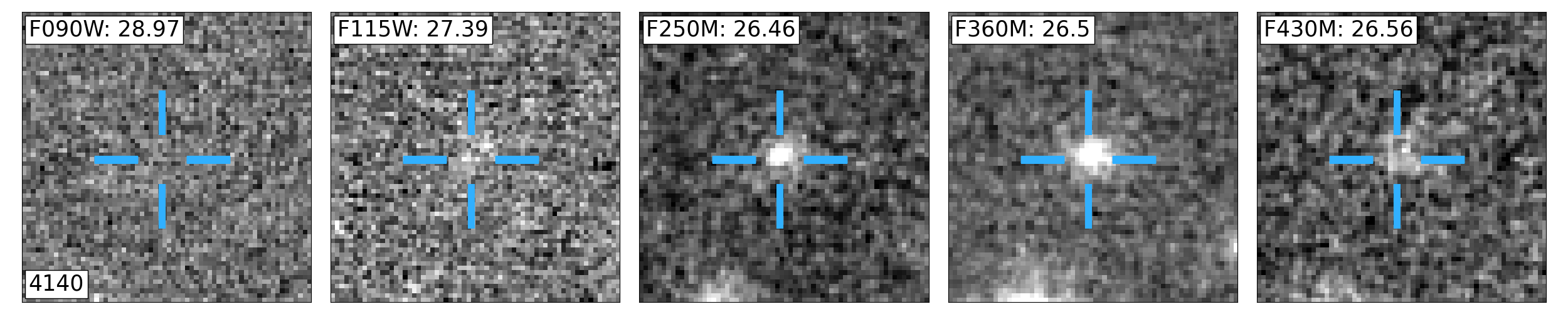}
\includegraphics[width=\linewidth]{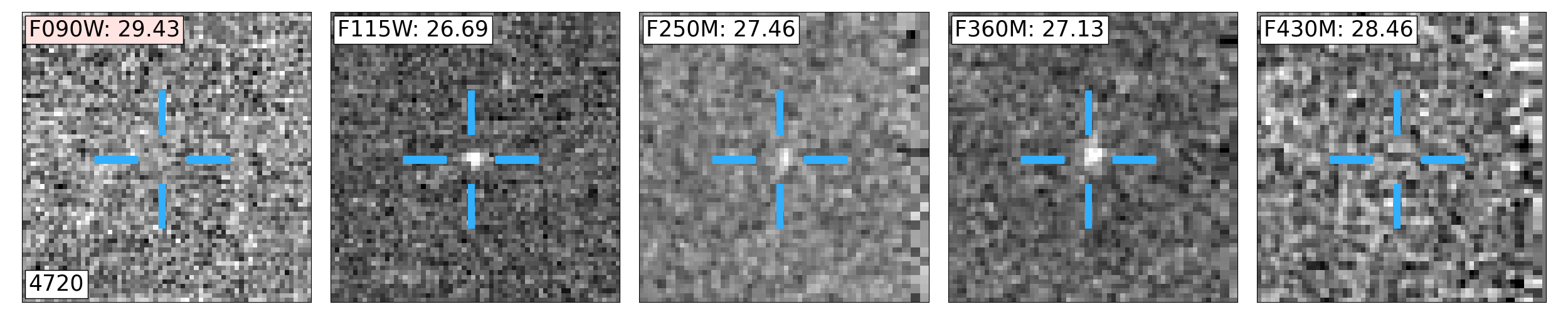}
\includegraphics[width=\linewidth]{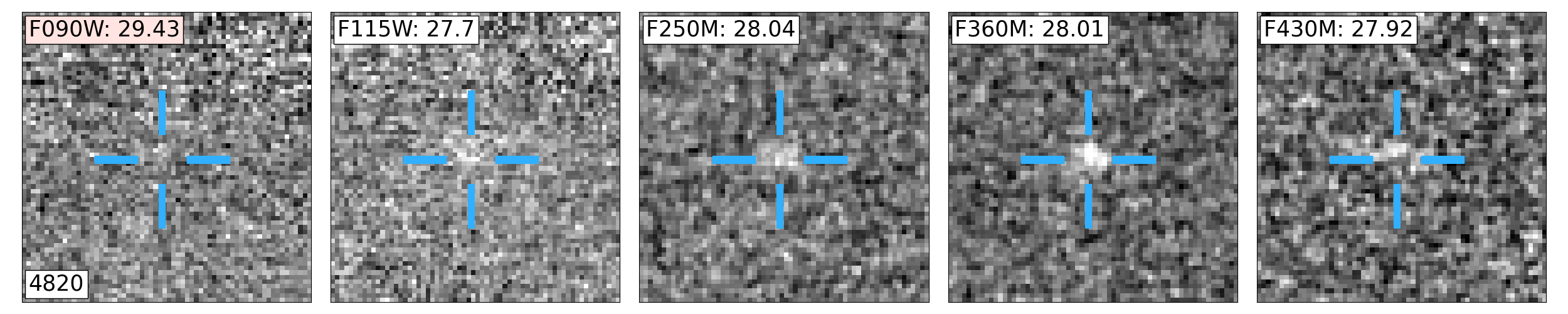}
\includegraphics[width=\linewidth]{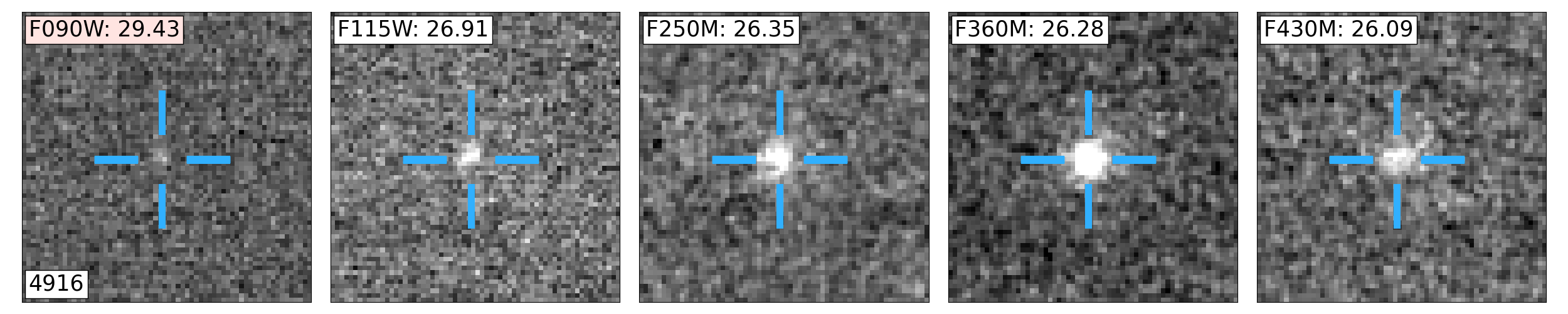}
\includegraphics[width=\linewidth]{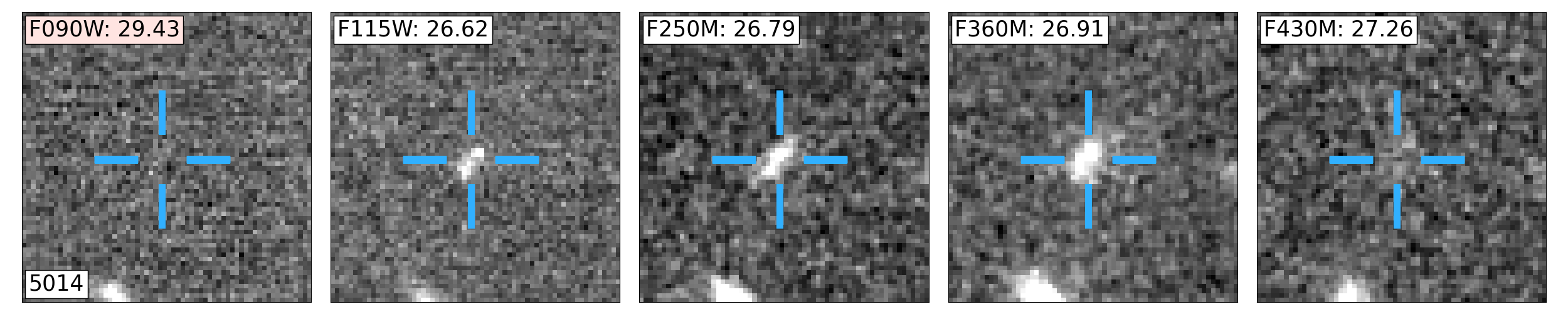}
\includegraphics[width=\linewidth]{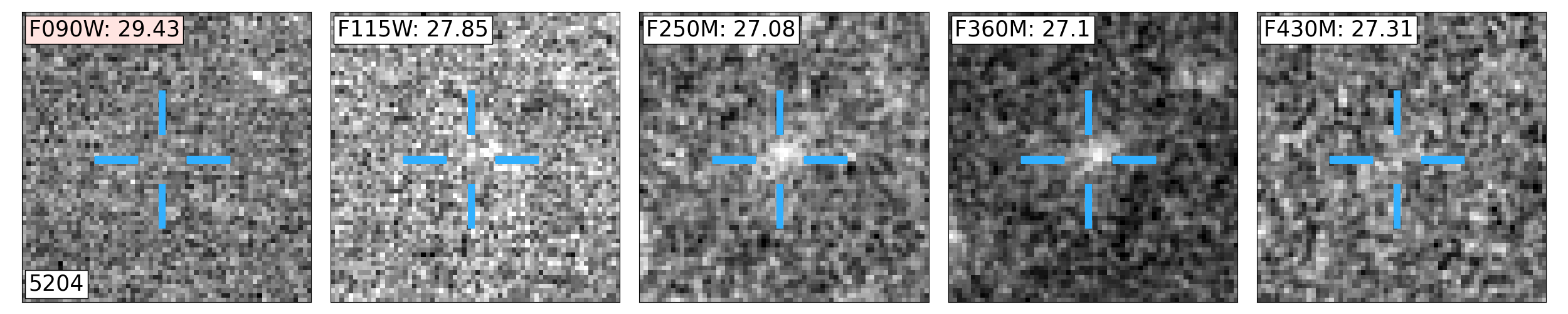}
\caption{See Figure~\ref{fig:cut1}.}
\end{figure}

\begin{figure}
\centering
\includegraphics[width=\linewidth]{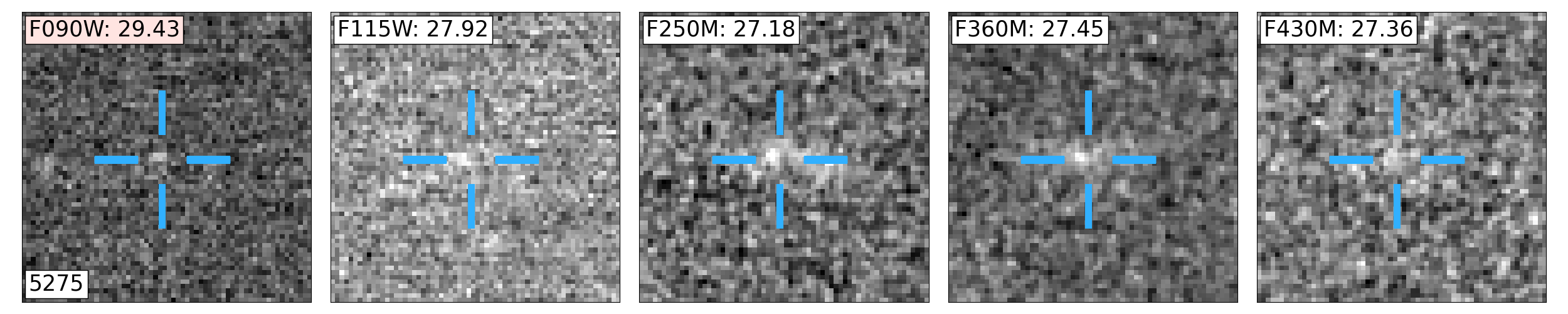}
\includegraphics[width=\linewidth]{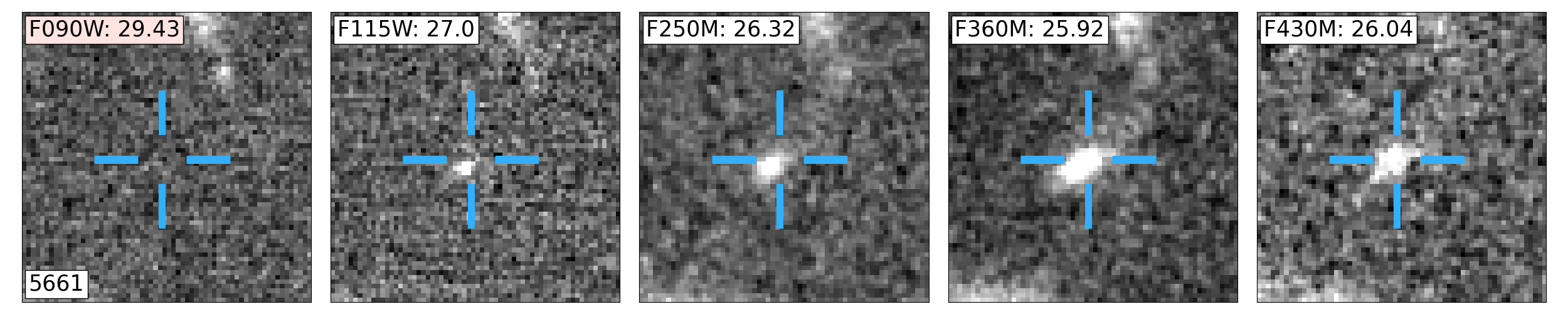}
\includegraphics[width=\linewidth]{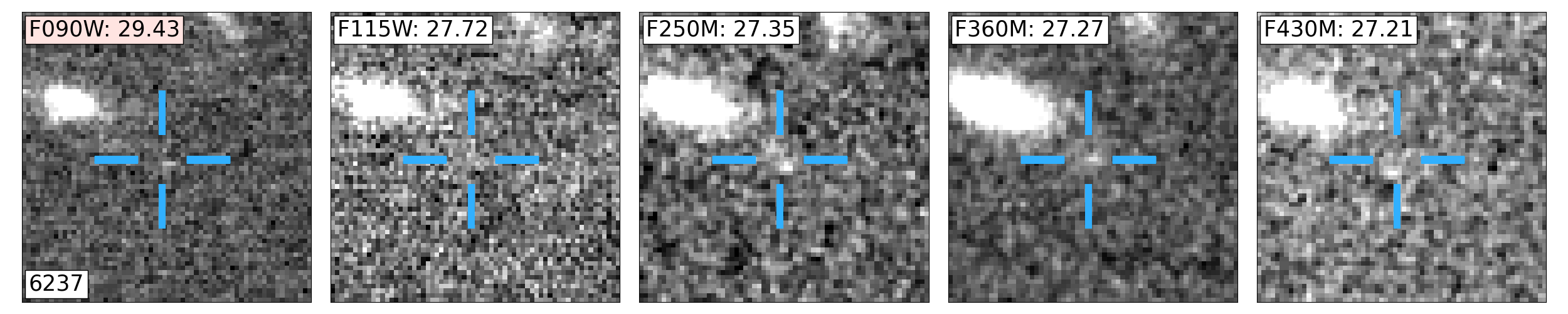}
\includegraphics[width=\linewidth]{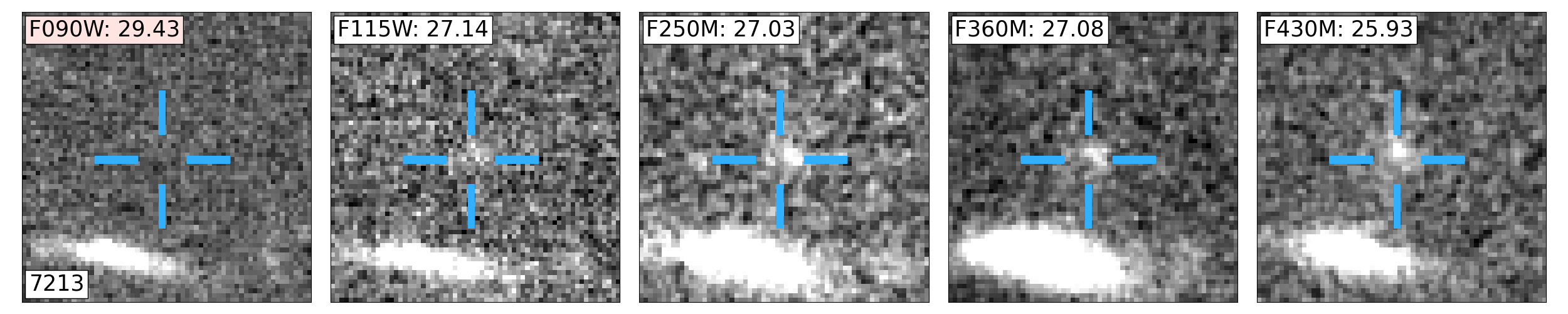}
\includegraphics[width=\linewidth]{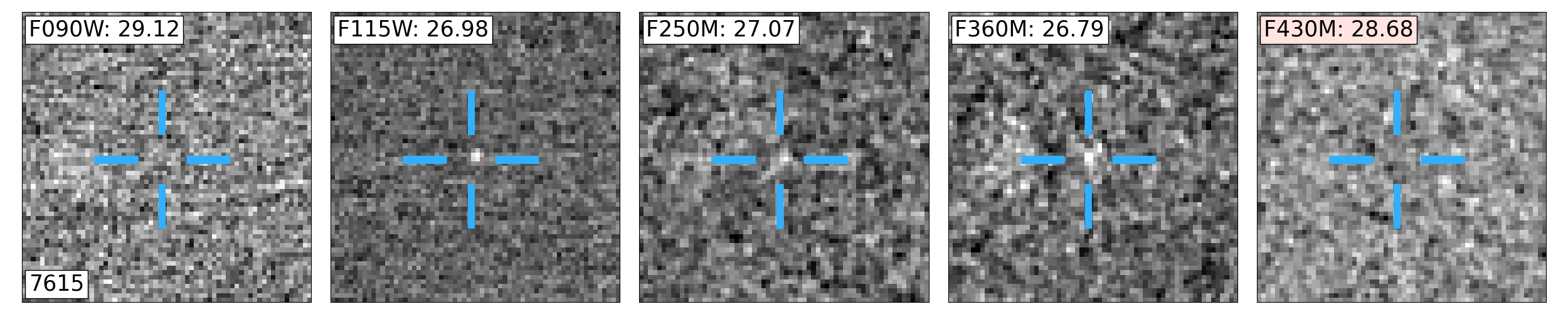}
\includegraphics[width=\linewidth]{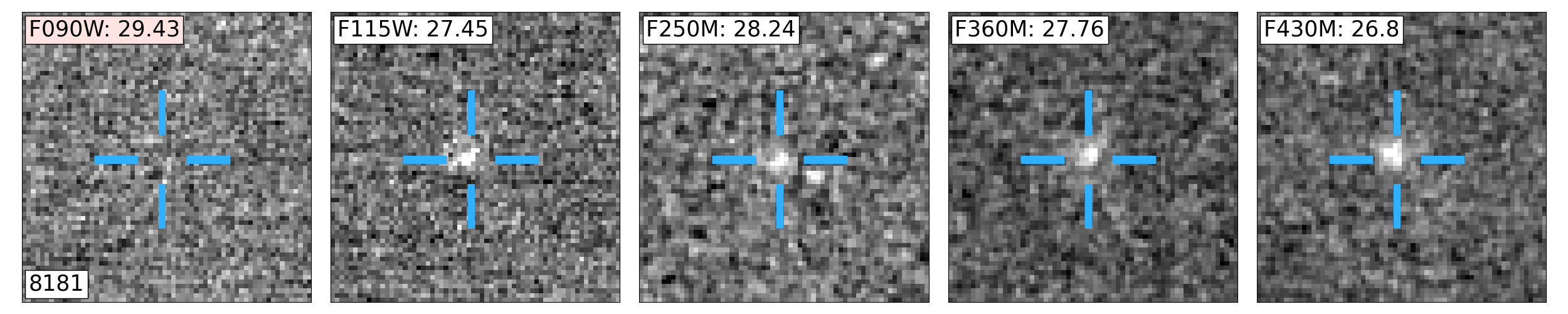}
\caption{See Figure~\ref{fig:cut1}.}
\end{figure}

\begin{figure}
\centering
\includegraphics[width=\linewidth]{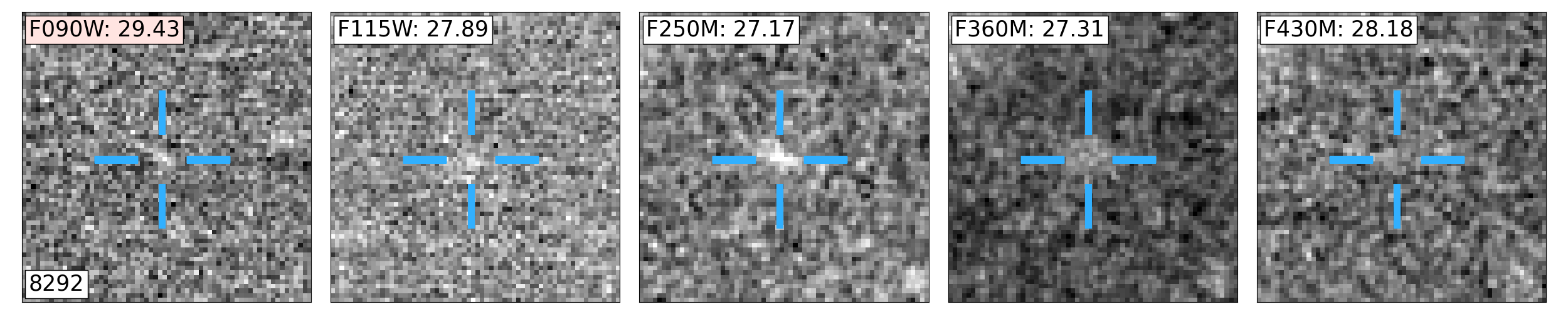}
\includegraphics[width=\linewidth]{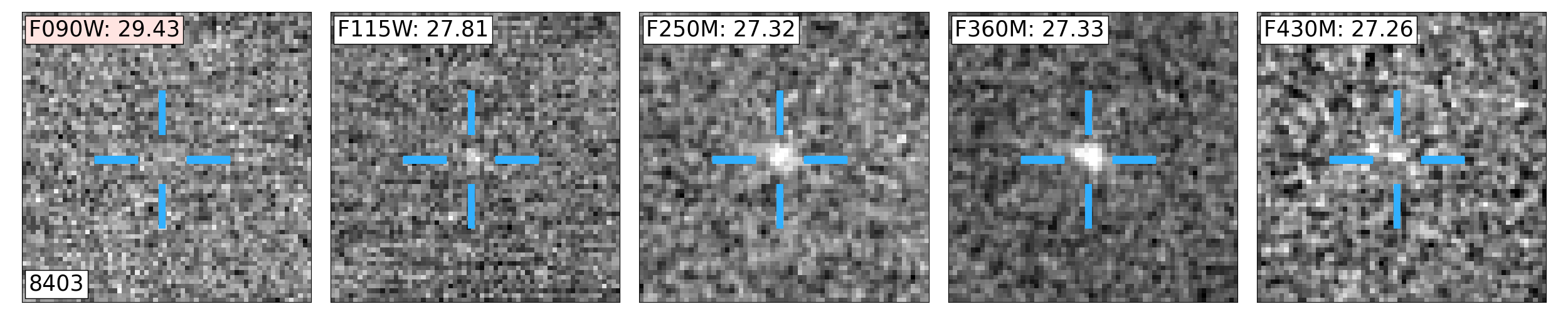}
\includegraphics[width=\linewidth]{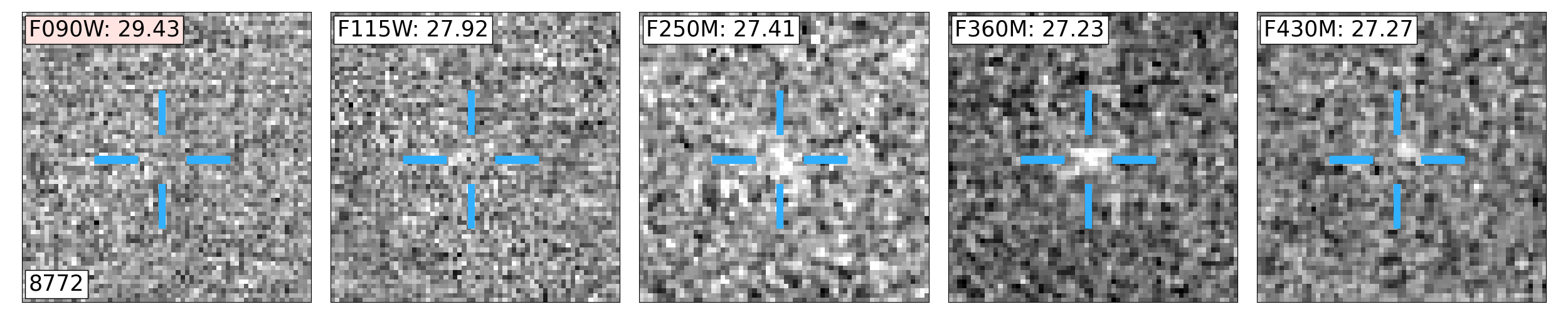}
\includegraphics[width=\linewidth]{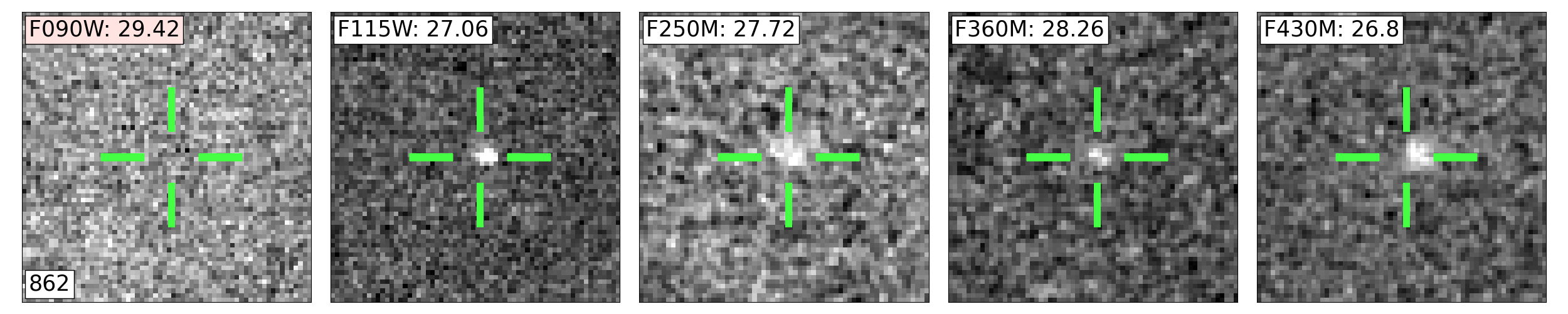}
\includegraphics[width=\linewidth]{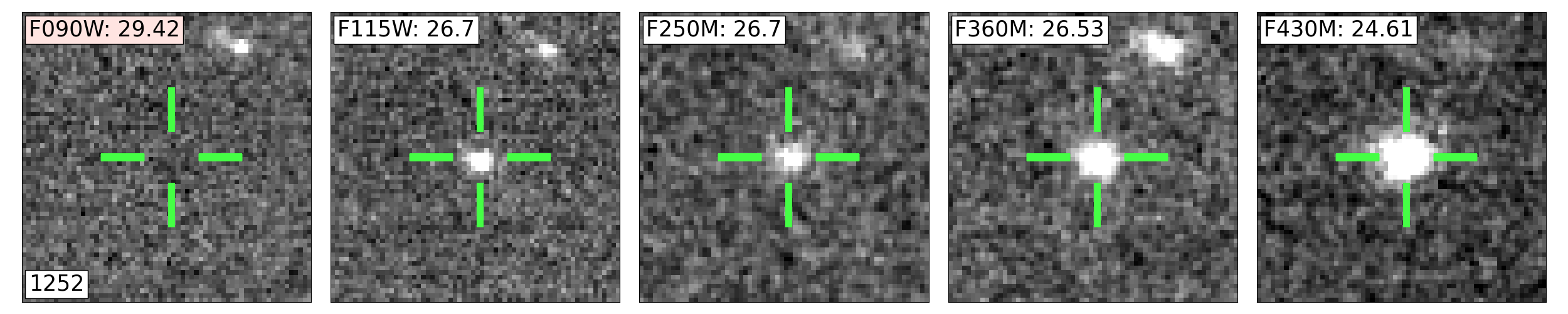}
\includegraphics[width=\linewidth]{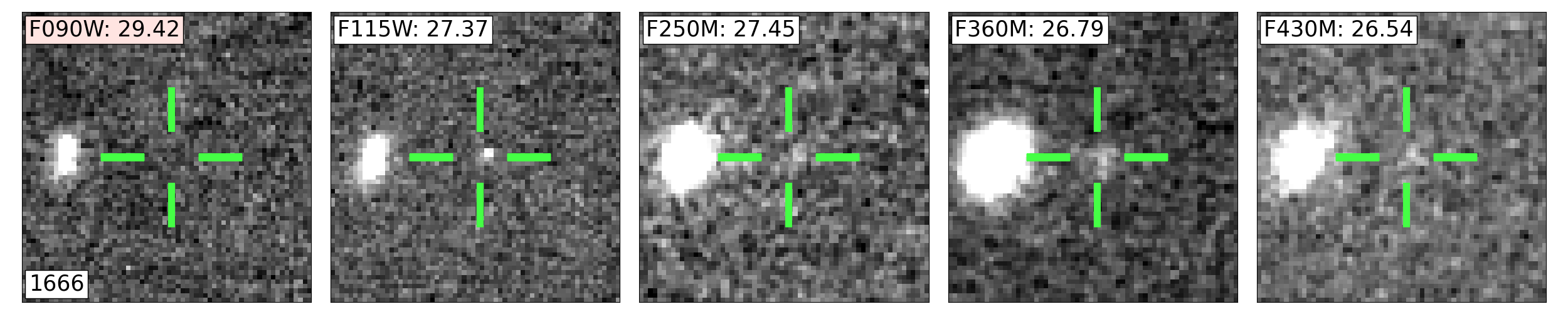}
\caption{See Figure~\ref{fig:cut1}.}
\end{figure}
\begin{figure}
\centering
\includegraphics[width=\linewidth]{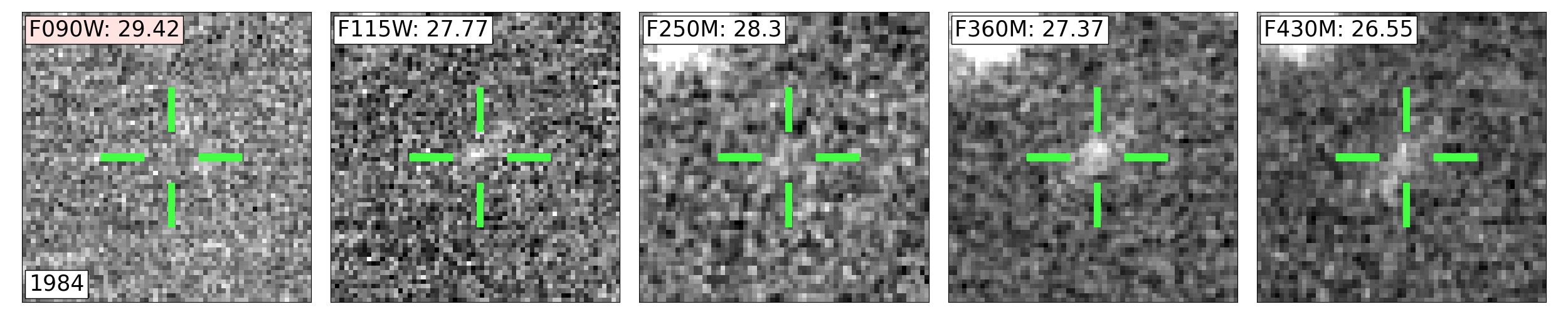}
\includegraphics[width=\linewidth]{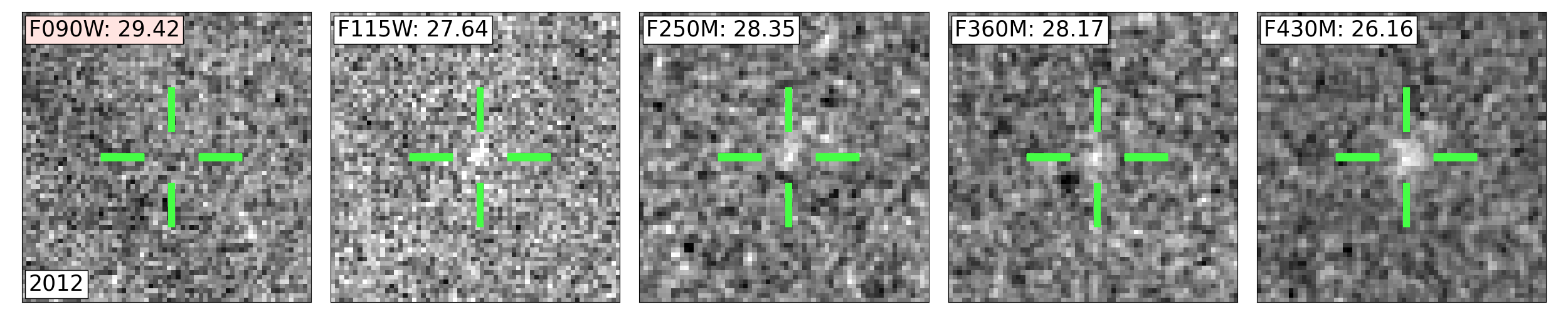}
\includegraphics[width=\linewidth]{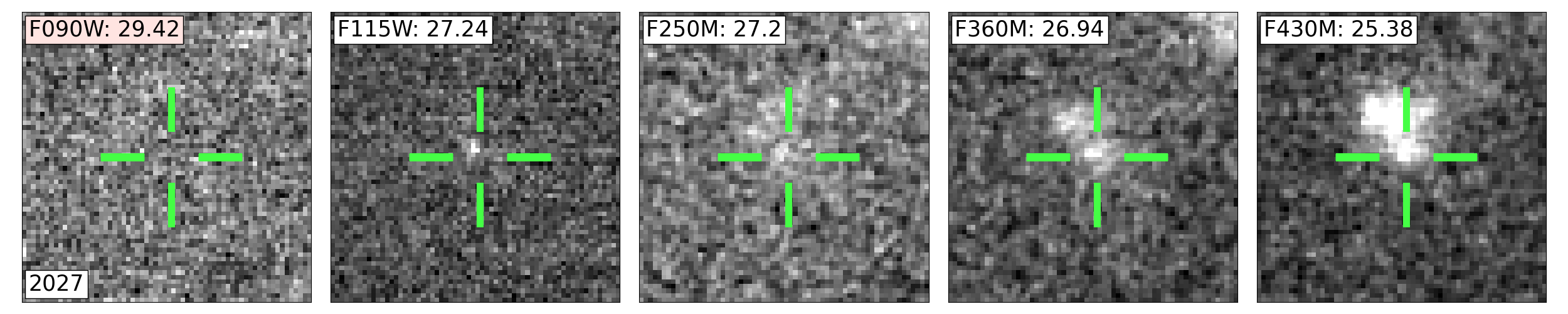}
\includegraphics[width=\linewidth]{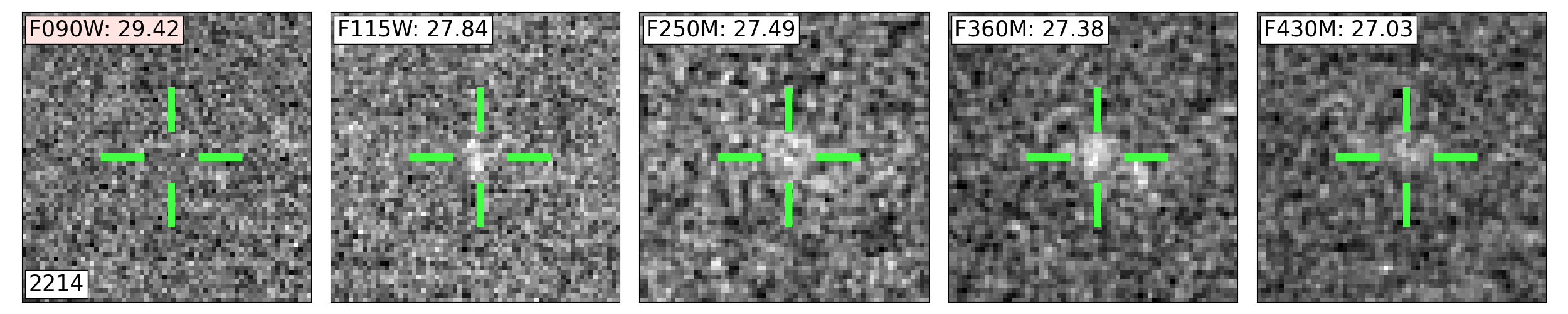}
\includegraphics[width=\linewidth]{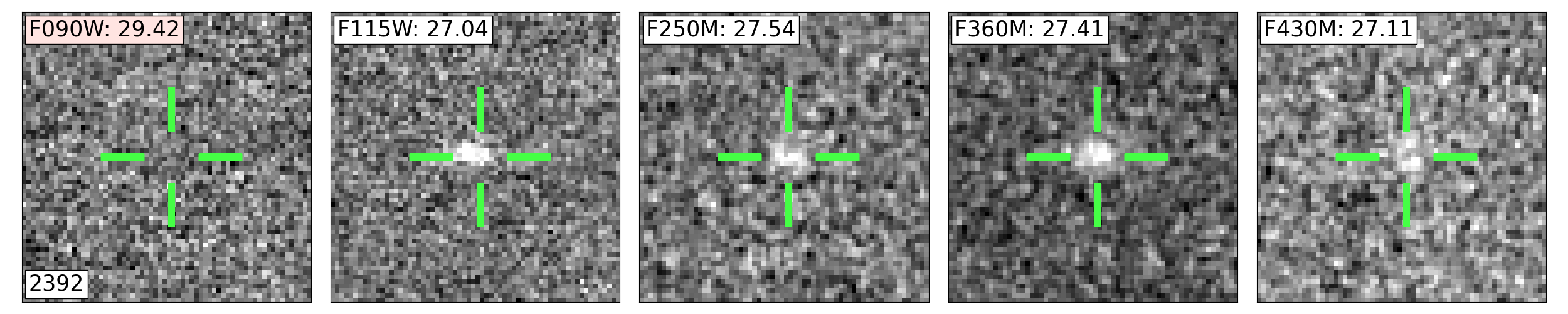}
\includegraphics[width=\linewidth]{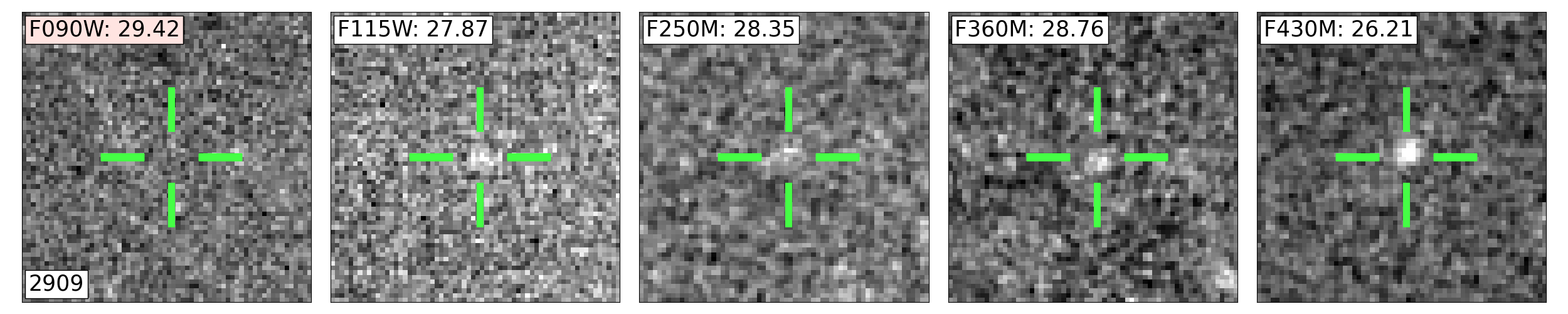}
\caption{See Figure~\ref{fig:cut1}.}
\end{figure}
\begin{figure}
\centering
\includegraphics[width=\linewidth]{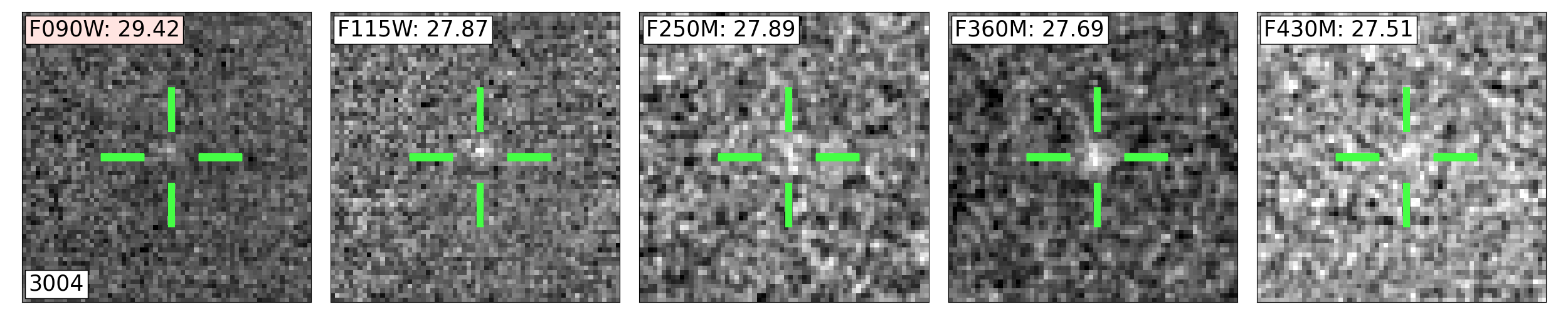}
\includegraphics[width=\linewidth]{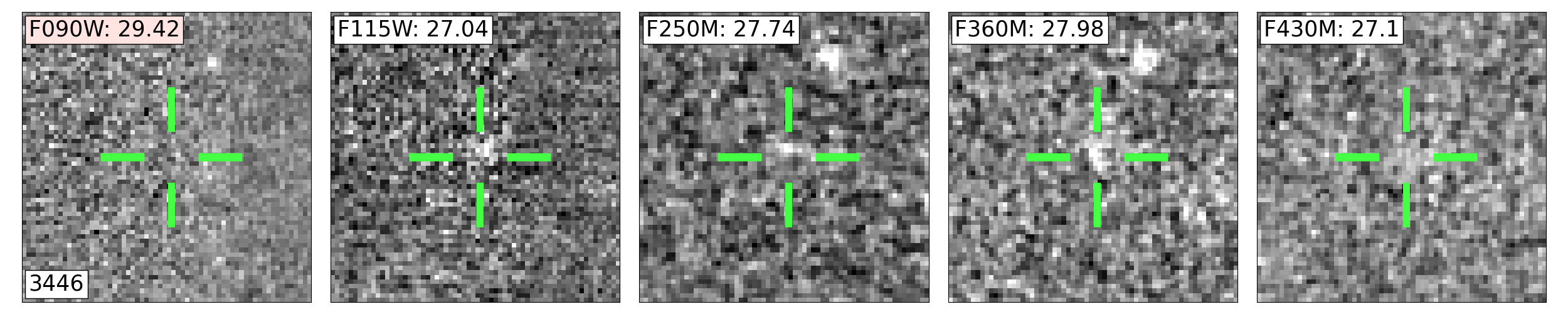}
\includegraphics[width=\linewidth]{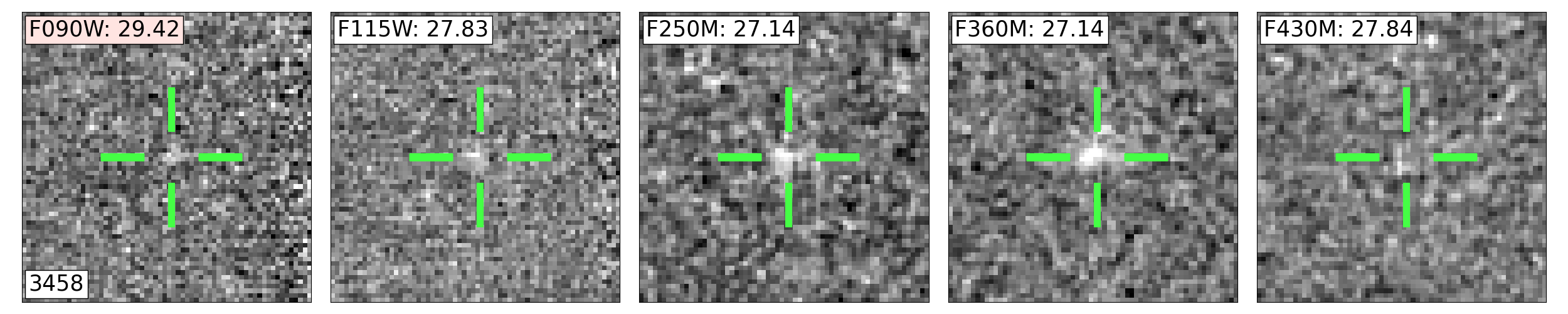}
\includegraphics[width=\linewidth]{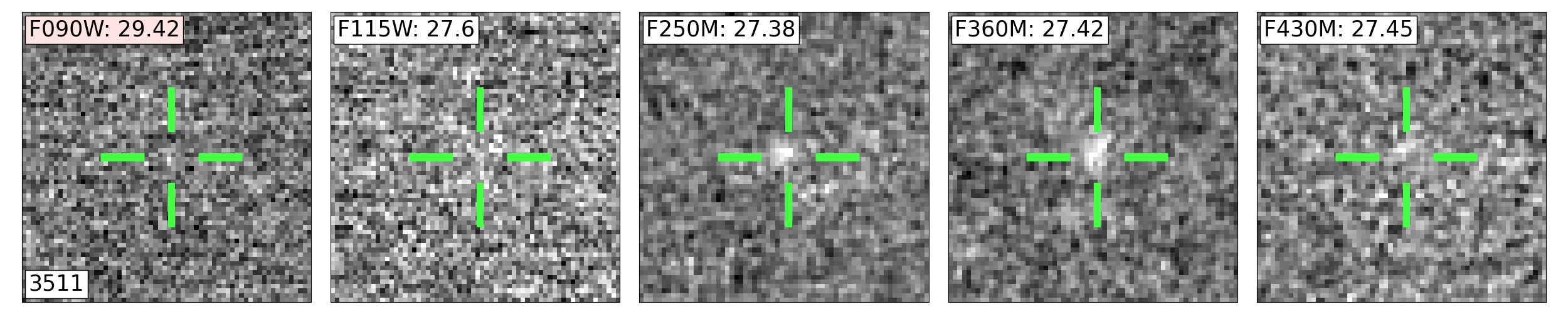}
\includegraphics[width=\linewidth]{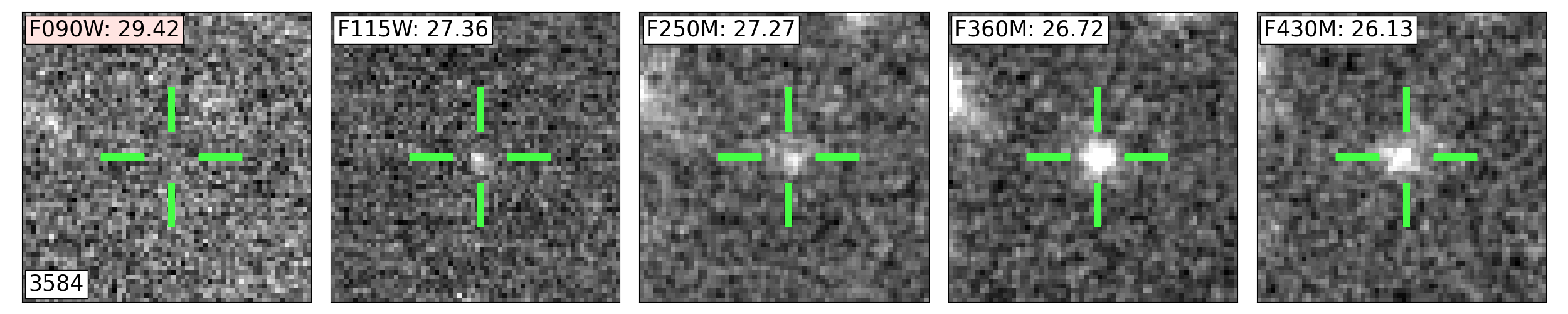}
\includegraphics[width=\linewidth]{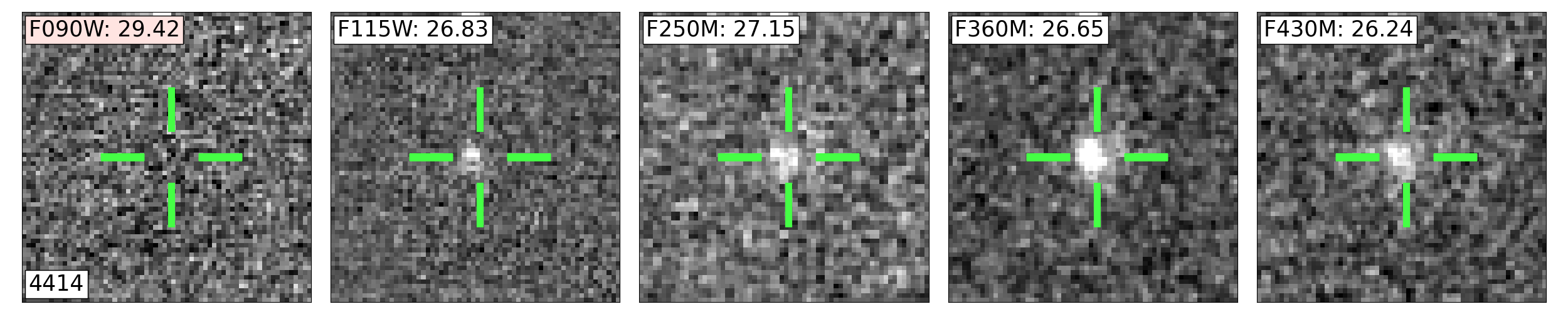}
\caption{See Figure~\ref{fig:cut1}.}
\end{figure}
\begin{figure}
\centering
\includegraphics[width=\linewidth]{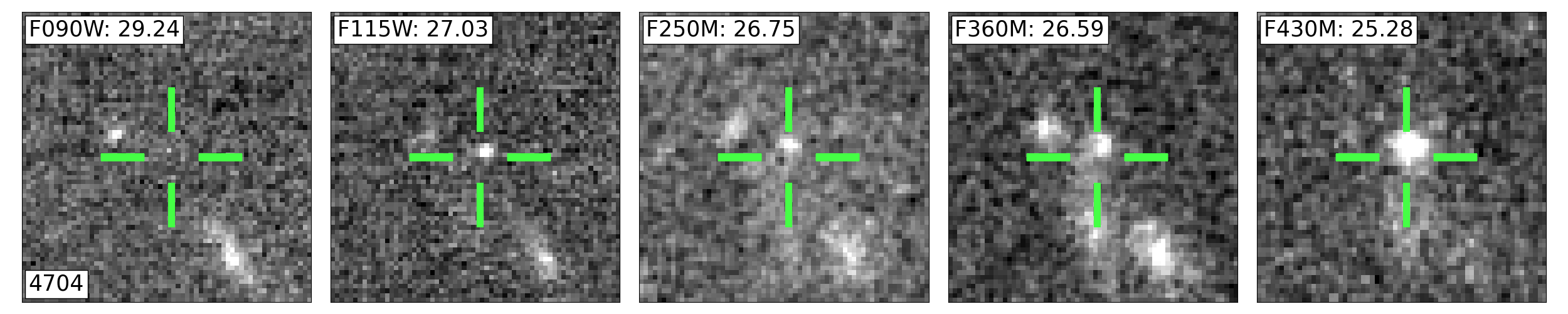}
\includegraphics[width=\linewidth]{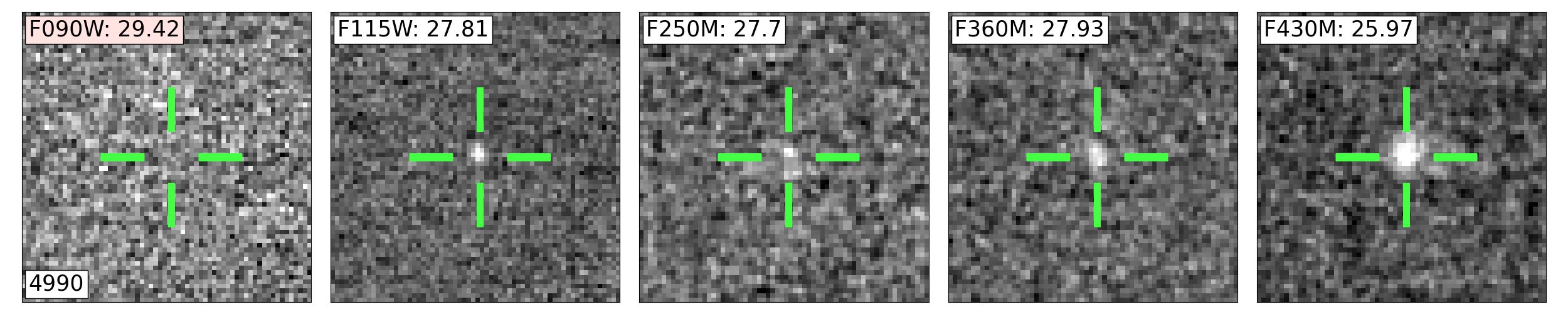}
\includegraphics[width=\linewidth]{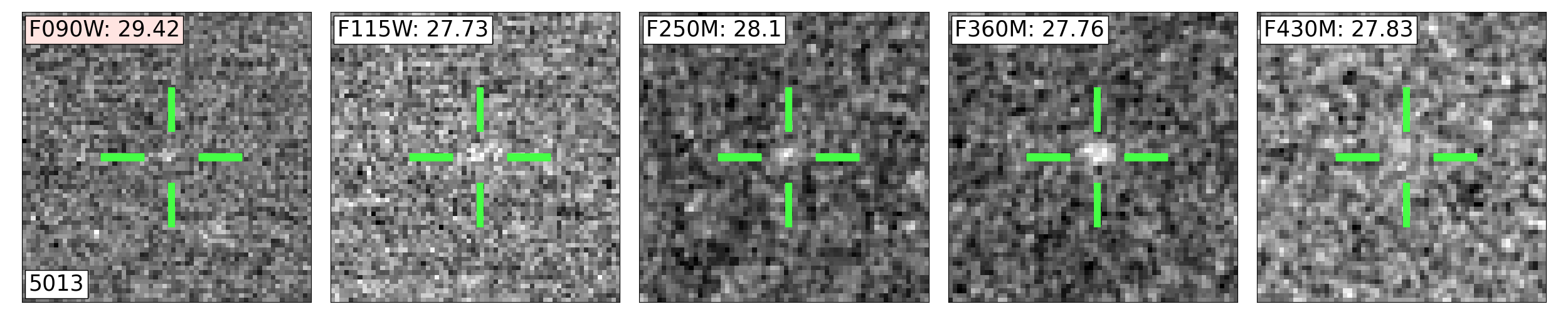}
\includegraphics[width=\linewidth]{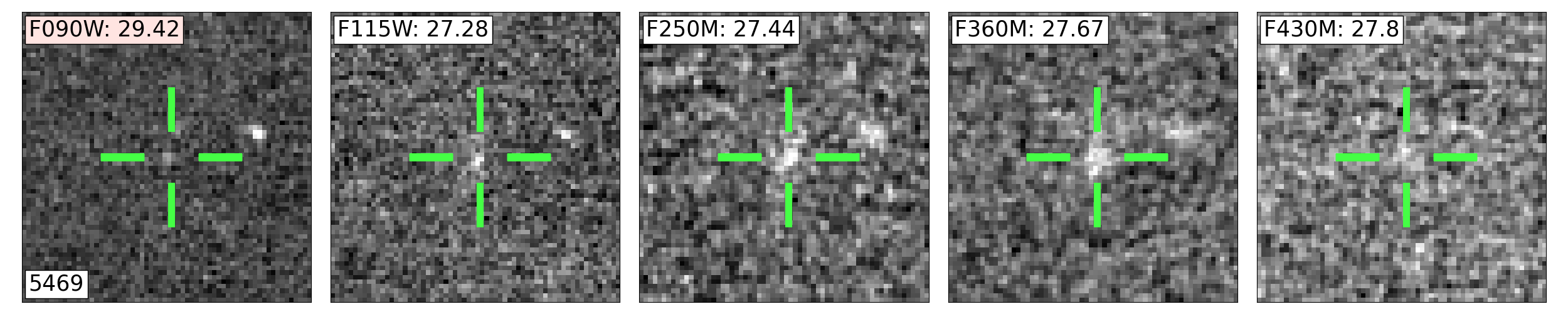}
\includegraphics[width=\linewidth]{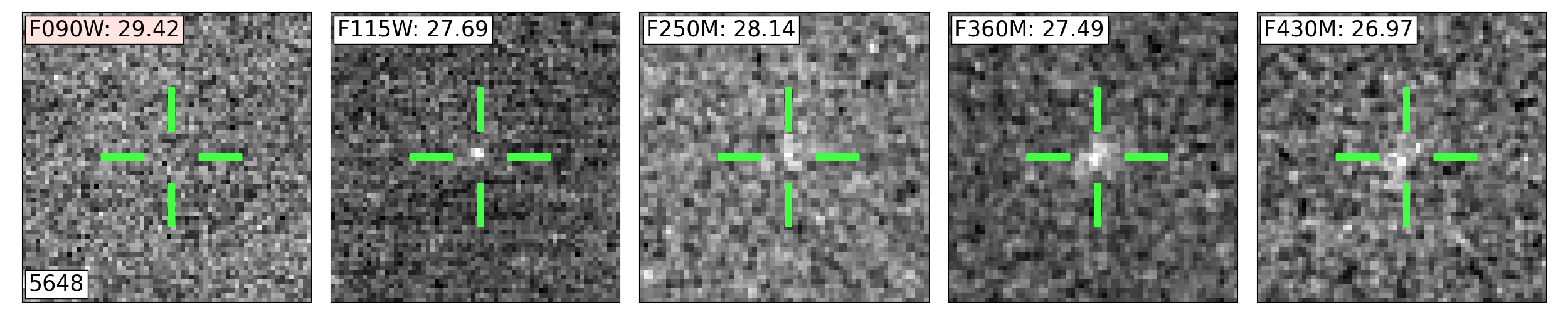}
\includegraphics[width=\linewidth]{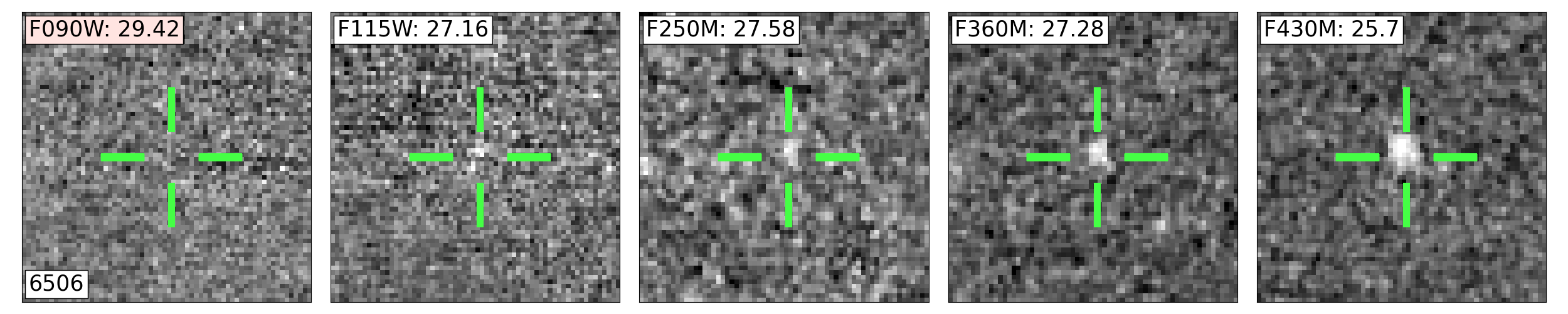}
\caption{See Figure~\ref{fig:cut1}.}
\end{figure}
\begin{figure}
\centering
\includegraphics[width=\linewidth]{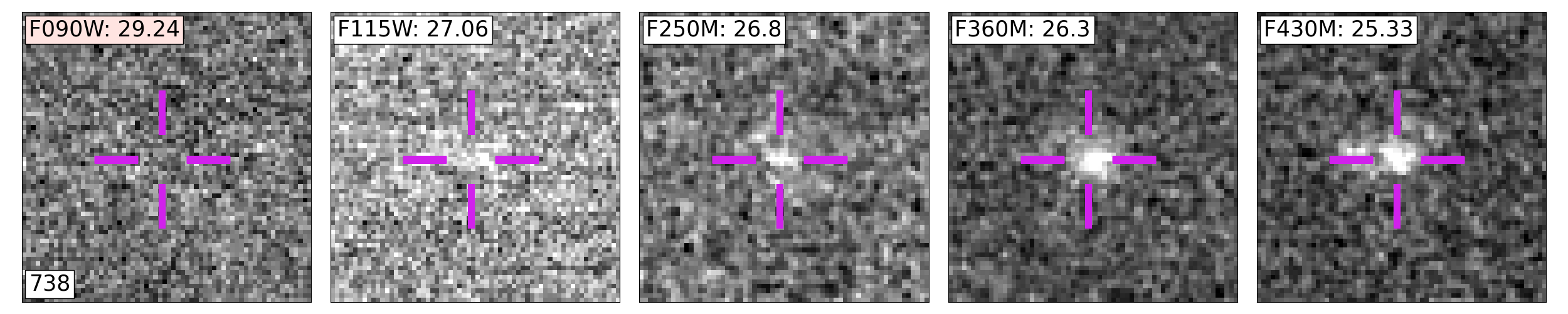}
\includegraphics[width=\linewidth]{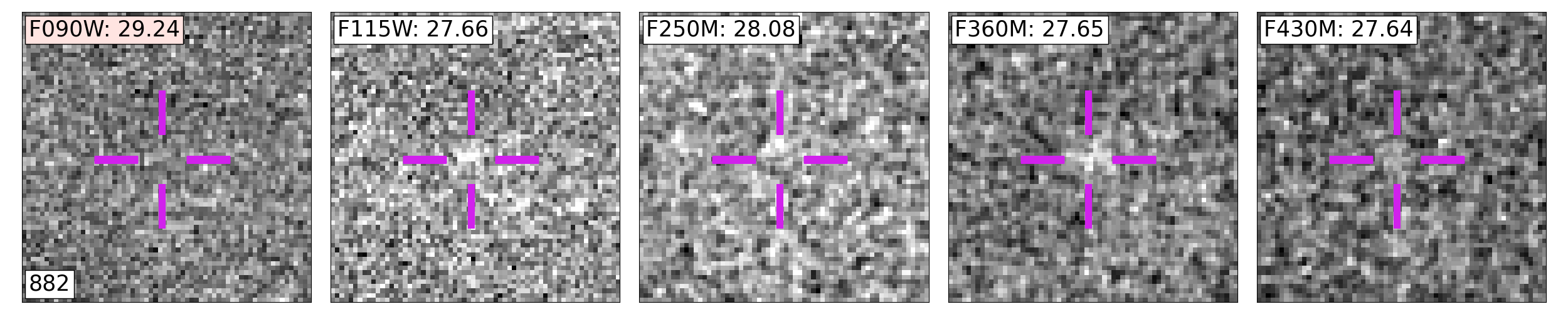}
\includegraphics[width=\linewidth]{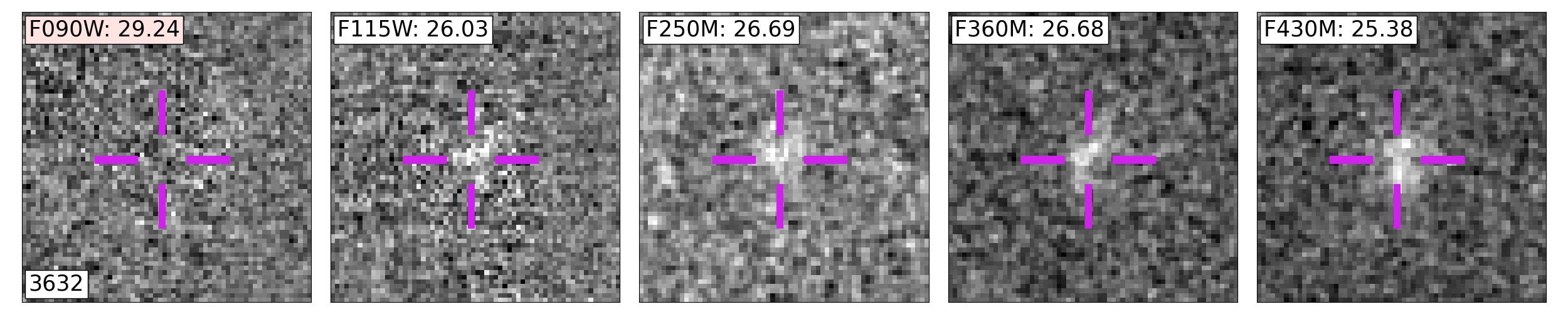}
\includegraphics[width=\linewidth]{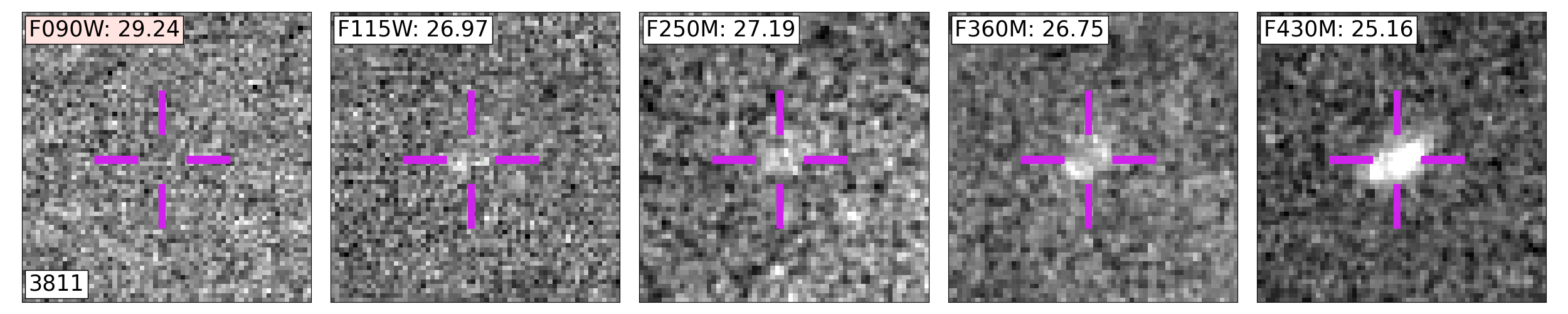}
\includegraphics[width=\linewidth]{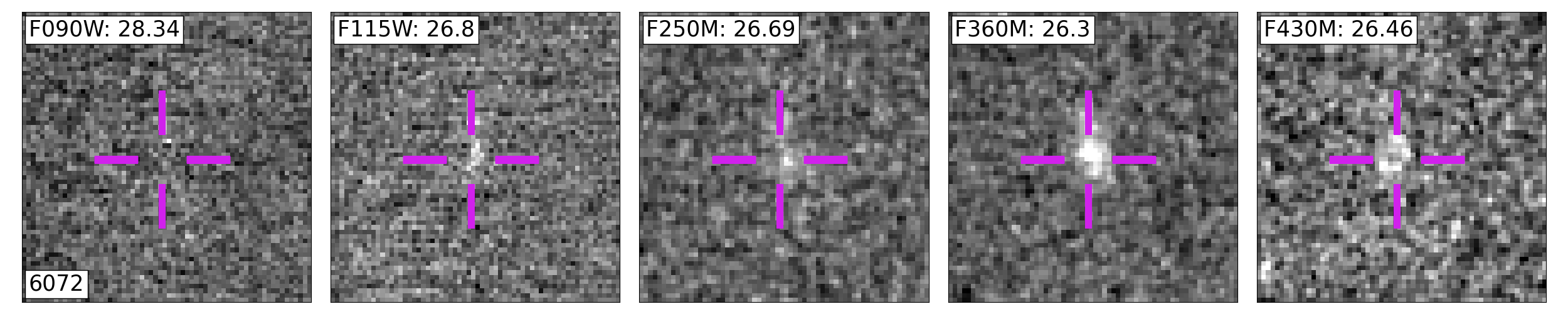}
\includegraphics[width=\linewidth]{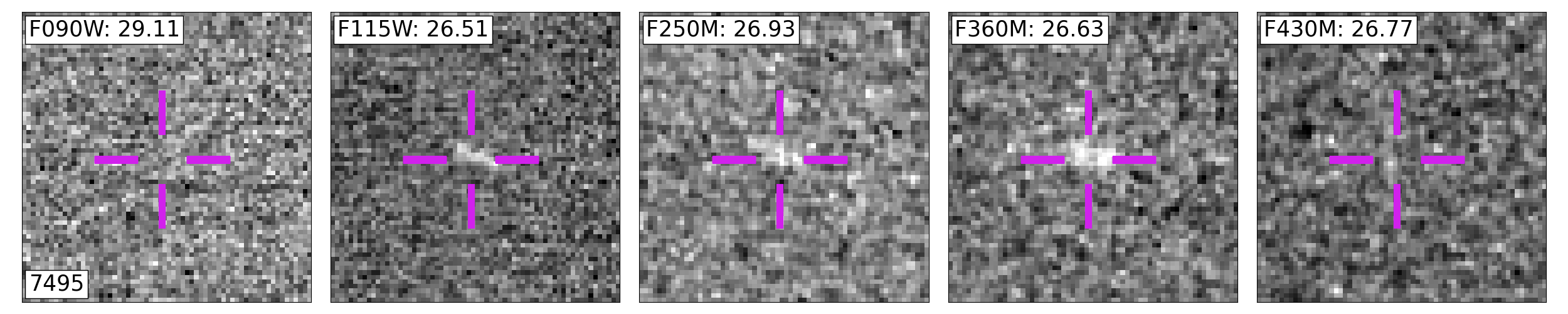}
\caption{See Figure~\ref{fig:cut1}.}
\end{figure}

\end{document}